\documentclass{article}[12pt]

\usepackage{graphicx}
\usepackage{tabu}
\usepackage{epsfig}
\usepackage{mathtools}
\usepackage{float}
\usepackage{algorithm}
\usepackage[noend]{algpseudocode}
\usepackage{amsmath}
\usepackage[export]{adjustbox}
\usepackage{subcaption}
\usepackage{tikz}
\usepackage{amsmath}

\newcommand{\expt}[1]{\langle#1\rangle}

\usepackage{amssymb}
\usepackage{authblk}

 \title{A General Model Validation and Testing Tool}

 \author{Kevin Vanslette, Tony Tohme, and Kamal Youcef-Toumi }
 \affil{Department of Mechanical Engineering\\ Massachusetts Institute of Technology, Cambridge MA 02139, USA}
 \date{\today}

\begin{document}
\maketitle
\begin{abstract}
We construct and propose the ``Bayesian Validation Metric" (BVM) as a general model validation and testing tool. We find the BVM to be capable of representing all of the standard validation metrics (square error, reliability, probability of agreement, frequentist, area, probability density comparison, statistical hypothesis testing, and Bayesian model testing) as special cases and find that it can be used to improve, generalize, or further quantify their uncertainties. Thus, the BVM allows us to assess the similarities and differences between existing validation metrics in a new light.   

The BVM has the capacity to allow users to invent and select models according to novel validation requirements. We formulate and test a few novel compound validation metrics that improve upon other validation metrics in the literature. Further, we construct the BVM Ratio for the purpose of quantifying model selection under user defined definitions of agreement in the presence or absence of uncertainty. This construction generalizes the Bayesian model testing framework.
\end{abstract}



 


%

%
%
%







	\section{Introduction}

	Pivotal to the scientific and engineering process is the testing of hypothesis and the validation of mathematical and computational models. Without a validation procedure, there is no reason to believe that a model, which has been designed to serve as a convenient representation of physical and/or human-made processes, has fulfilled its functional purpose. Thus, validation is the result of the positive justification of a model's representation of relevant features in the real world \cite{Oberkampf2004,Sornette,Trucano}. 
	
	We are interested in studying the validation of multivariate computational models that represent uncertain situations and/or data. It is understood that complete certainty is a special case of uncertainty.
	The uncertainty in a model or data set may originate from stochasticity, model parameter and input data uncertainty, measurement uncertainty, or other possible aleatoric or epistemic sources of uncertainty.  Each of the following data modeling schemes may include quantifiable amounts of uncertainty (or certainty) that we would like to validate on the basis of a set of validation data: neural networks and AI models, machine learning models, Gaussian Process Regression models \cite{KOH}, polynomial chaos and other surrogate models \cite{Spectral,Dakota,UQtk,NumChallenges}, spatial and time series stochastic models, physics based models (usually solutions to differential equations), engineering based models (which are sufficiently abstracted physics based models), Monte Carlo simulation models \cite{MCMC,Metro}, and more. Model output uncertainties may be quantified through uncertainty propagation techniques (that may or may not include verification, calibration, and validation) \cite{KOH,Dakota,UQtk,NumChallenges,Shankar,Sankararaman,Mahadevan,Li,Roy,UQlab,MUQ}. 	
	
	There exist several validation metrics. Each metric is designed to compare features of a model-data pair to quantify validation: square error compares the difference in the data and model values in a point to point or interval fashion \cite{epsilonmetric}, the reliability metric \cite{Rebba} and the probability of agreement \cite{Stevens} compare continuous model outputs and data expectation values (the model reliability metric was extended past expectation values in \cite{Sankararaman2}), the  frequentist validation metric \cite{Oberkampf,OberkampfAIAA} and statistical hypothesis testing compare data and model test statistics, the area metric compares the cumulative distribution of the model to the estimated cumulative distribution of the data \cite{Ferson,Roy,Ling,Li2,Wu,Zhao}, probability density function (pdf) comparison metrics such as the KL Divergence that represent ``closeness" between pdfs, and Bayesian model testing compares the posterior probability that each model would correctly output the observed data \cite{Sankararaman,Mahadevan,Sivia,Placek, Gelfand,Geweke, Zhang}. A detailed review of the majority of these metrics may be found in \cite{Liu2,Maupin,Lee,Mullins} and the references therein. In particular, \cite{Maupin} is an up to date review that considers many validation metrics in the cases of data and model certainty, data uncertainty and model certainty, and data and model uncertainty. Further, they comment on the importance of model validation while admitting its contextual difficulties because often validation is ``strongly application, quantity of interest, and modeler, dependent". Our article addresses some of these contextual difficulties by taking a bottom up approach.  

	To assist the comparison of the positive and negative aspects of (most) the above validation metrics, reference \cite{Liu2} outlines six ``desirable validation criteria" that a validation metric might have (they extend \cite{Oberkampf,Ferson}). One conclusion from \cite{Liu2} is that none of the available metrics simultaneously satisfy all six desirable validation criteria. We summarize the most important features of the desirable validation criteria with the following validation criterion: 
	\begin{enumerate}
	\item A validation metric should be a statistically quantified quantitative measure (as opposed to a qualitative measure) of the \emph{agreement} between (general) model predictions and data pairs, in the presence or absence of uncertainty.
	\end{enumerate}

	The desire for objectivity, ``that a metric will produce the same assessment for every analyst independent of their individual preferences" \cite{Liu2}, is difficult to satisfy because there are no rules in place to guide a modeler toward selecting one validation metric over another. For this reason the individual might simply choose a metric based on their preferences, or worse, be tempted to base their decision on which validation metric gives them the most favorable evaluation. Given individuals may choose different validation metrics for the same model-data pair, it is possible for individuals to impose accuracy requirements that are incompatible with one another and arrive at different conclusions regarding the validity of the same model-data pair. As the final goal is objectivity, when possible, a map between the accuracy requirements should be constructed such that the validation metrics yield consistent evaluations of model-data validity when applicable.
		
	Further, Liu et al. \cite{Liu2} suggest that there is no agreed upon unified model-data comparison function. Even including the results of this article, we expect this statement to hold as it is extremely difficult to guess the prior information about the utility of a model an analyst may be required to include in the validation of a model-data pair. For instance given arbitrary data  ``What features of the data are relevant to capture with a model?", ``Of these features, are some more relevant than others?", and  ``What accuracy is required for the model to be valid?". \emph{Agreement} and \emph{validation} are ultimately human-made concepts designed for the purpose of expressing that ``in general, not every feature or statistic between a model-data pair need to be equal to conserve the utility of the model". For some model-data pairs, all that may be required is that the model and data averages closely match within uncertainty, while for others, one may require that the model can accurately reproduce the probability distribution of a data set as a whole (as one would do to physically model a noisy measurement device). Given the wide variety of data and the large number of different inferences (and thus models and hypotheses) that one may be interested in drawing from a given data source, i.e. the context of the model-data pair, we do not expect any single set of comparison functions, statistics, or values to be equally relevant and maximally useful for all possible model-data contexts. This, however, does not stop us from quantifying the validity of a model-data pair given any arbitrary comparison function and with any arbitrary definition of agreement.

In this article, we construct and propose the ``Bayesian Validation Metric" (BVM) as a general model validation and testing tool. We design the BVM to adhere to the desired validation criterion (1.) by using ``four BVM inputs": the model and data comparison values, the model output and data pdfs, the comparison value function, and the agreement function. The comparison value function is a function of model output and data comparison values that provides the desired quantitative comparison measure, e.g. square difference. Using the model output pdf and the data pdf, the value of the comparison value function is statistically quantified. In turn, the agreement function provides an accept/reject rule and effectively wraps the previous three BVM inputs together to give the BVM. From this, the BVM outputs ``the probability the model and data agree", where \emph{agreement} is a user defined Boolean function that meets, or does not meet, accuracy requirements between model and data comparison values. Thus, the BVM meets the desired validation criterion (1.) for arbitrary comparison value functions, arbitrary definitions of agreement, and in principle for arbitrary data types such as integers, vectors, tensors, strings, pictures, or others.

	
	The BVM can be used to represent all of the aforementioned validation metrics as special cases. This allows us to compare and contrast the validation metrics from the literature in a new light. We find the conditions under which several of the current validation metrics are effectively equal to one another, which improves the objectivity of the current validation procedure. In brief we find that the frequentist metric (using natural definitions of agreement) is equal to the reliability metric and the probabilities from Bayesian model testing are equal to the probabilities of the improved model reliability metric \cite{Sankararaman2} when one demands exact equality of the (uncertain) model-data comparison values. Because probability can represent both certain and uncertain situations, so can the BVM. Thus, these ``special case" metrics can be generalized to quantify certain or uncertain cases, and even be combined into more complex validation requirements using the BVM framework. Thus, the BVM provides a standardized framework to improve, generalize, or further quantify these validation metrics. 
	
	\begin{figure} [H]		
\begin{centering}

  \includegraphics[scale=0.85]{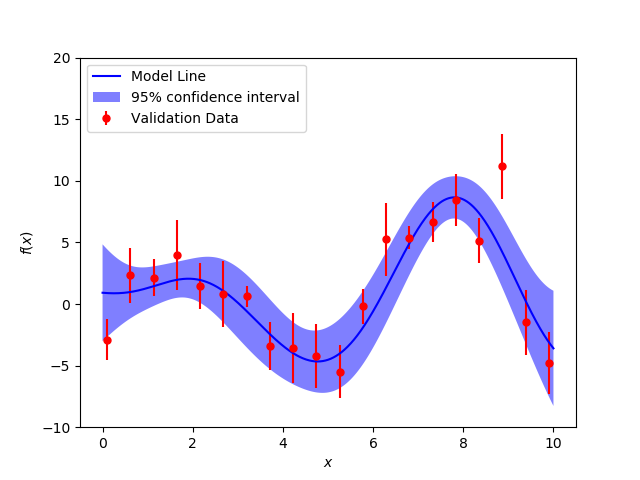}
\caption{\label{fig:tonyplot} This is the depiction of a common model validation scenario. The model line is trained on noisy data (not depicted in the figure) and is to be compared to a set of validation data. As both the model line and the data are uncertain in general, any quantitative measure (i.e. the comparison function) between these comparison values inherits this uncertainty. Thus, any accept/reject rule on the basis of these uncertain comparison function values is uncertain as well. A visual inspection of this graph seems to indicate, up to statistical fluctuation, that the comparison values of the model and data more or less (or probably) agree, but this intuitive measure has yet to be quantified. Graphic adapted from \cite{scikit}.}
\end{centering}
\end{figure}
\noindent 

By constructing the ``BVM ratio", we generalize the Bayesian model testing framework \cite{Sivia}. 
In the Bayesian model testing framework, one constructs the Bayes ratio to rank models according to the ratio of their posterior probabilities given the data. We show that these posterior probabilities are equal to a special case of the BVM under the definition of agreement that requires these uncertain model outputs and data to match exactly. Thus, nothing prevents us from extending the logic used in the Bayesian model testing framework to our framework and we construct the BVM ratio for the purpose of model selection under arbitrary definitions of agreement, i.e. for arbitrary validation scenarios. 

	The remainder of the article continues as follows. In Section 2 we construct the BVM by following our validation criterion. Through some edge cases, we show that the BVM satisfies both the six desirable validation criteria from \cite{Liu2} as well as our validation criterion (1.) in Section 3. In Section 4, we summarize the results derived in Appendix A. There, we incorporate all of the above standard validation metrics as special cases of the BVM, draw relationships between several of the validation metrics, provide improvements and generalizations to these metrics as is suggested by the functional form of the BVM, and construct the BVM ratio.
	In Section 5 we represent three novel validation metrics using the BVM and compare them to similar metrics from the literature. Due to the proven capacity of our framework, we consider the BVM to be a general model validation and testing tool.

	\section{The Bayesian Validation Metric} 
	For the remainder of the article we will use the following notation and language. We will let $\hat{y}$ denote the output of a model, $\hat{Y}$ a set of model outputs, $y$ a data point, and $Y$ a set of data points. The proposition $M$ essentially stands for ``the model" or ``coming from the model", and $D$ stands for ``the experiment" or ``coming from the experiment". We let $\hat{z}$ and $z$ represent the comparison quantities of interest, which pertain to the model and the data respectively. Further, we let the comparison quantities take general forms, such as multidimensional vectors, functions, or functionals (e.g. output values, expectation values, pdfs, ...), so we can represent any such pair of quantities we may wish to compare between the model and the data. When we refer to ``the four BVM inputs" we mean: the comparison values $(\hat{z},z)$, the model output and data pdf $\rho(\hat{z},z|M,D)$, the comparison value function $f(\hat{z},z)$, and the agreement function $B=B(f(\hat{z},z))$. The (denoted) integrals may be integrals or sums depending on the nature of the variable being summed or integrated over, which is to be understood from the discrete or continuous context of the inference at hand. The dot `` $\cdot$ " represents standard multiplication, which is mainly used to improve aesthetics.
	
		Performing uncertainty propagation through a model results in a model output probability (density) distribution $\rho(\hat{y}|M,D)$ that ultimately we would like to validate by comparing it to an uncertain validation data source $\rho(y|D)$, to see if they agree (as depicted in Figure 1). The immediate question is, however, ``What values do we want to \emph{compare} and what do we mean by \emph{agree}?". Given the wide variety of data and the large number of different inferences (and thus models and hypotheses) that one may be interested in drawing from a given data source, i.e. the context of the model-data pair, we do not expect any single set of comparison functions to be equally relevant and maximally useful for all possible model-data contexts. In light of this, we instead quantify the validity of a model-data pair given any arbitrary comparison value function and with any arbitrary definition of agreement.

	\subsection{Derivation}
	
	Here we will begin constructing the Bayesian Validation Metric (BVM). To capture the concept of what we might mean by \emph{agree}, we define $\hat{z}$ and $z$ to agree, $A$, when the Boolean expression, $B$, is true. Both $A$ and $B$ are defined by the modeler and their prior knowledge of the context of the model-data pair. Naturally then, the agreement function $B=B(f(\hat{z},z))=B(\hat{z},z)$ is some function or functional of a comparison value function $f(\hat{z},z)$.  
	
	Given the values of $\hat{z}$ and $z$ are known, i.e. certain, we quantify \emph{agreement} using a probability distribution that assigns certainty,
	\begin{eqnarray}
	p(A|\hat{z},M,z,D) =\Theta\Big(B(\hat{z},z)\Big).\label{agree}
	\end{eqnarray}
	The indicator function $\Theta\Big(B\Big)$ is defined to equal unity if $B$ evaluates to ``true" (i.e. ``agreeing") and equal to zero otherwise. Thus in the completely certain case, we are certain as to whether the model and data comparison values `agree, or do not agree, \emph{as defined by $B$ and the deterministic evaluation of $f(\hat{z},z)$}.\footnote{This binary yet probabilistic definition of agreement turns out to be completely satisfactory for our current purposes. As is briefly discussed later, the sharp boundaries of the indicator function can be smoothed out without employing fuzzy logic by allowing parameters in the Boolean function to themselves be uncertain and marginalized over. 
	} We will call $p(A|\hat{z},M,z,D)$ the ``agreement kernel". 
	
	Given that in general the comparison values are uncertain, and quantified by $\rho(\hat{z},z|M,D)$, the probability the comparison values agree, \emph{as defined by $B$ and $f(\hat{z},z)$}, is equal to,
	\begin{eqnarray}
	p(A|M,D)&=&\int_{\hat{z},z} p(A|\hat{z},M,z,D)\cdot\rho(\hat{z},z|M,D)\,d\hat{z}\,dz,\label{genagree}\\
	&\stackrel{\tiny{ind.}}{\longrightarrow}&\int_{\hat{z},z} \rho(\hat{z}|M,D)\cdot\Theta\Big(B(\hat{z},z)\Big) \cdot\rho(z|D) \,d\hat{z}\,dz \label{genagree2},
	\end{eqnarray}
	which is a marginalization over the spaces of $(\hat{z},z)$.\footnote{ Recall that the propositions in the probability distributions $\rho(z|D)$ and $\rho(\hat{z}|M,D)$ are completely arbitrary (in some cases requiring propagation from $\rho(y|D)$ and $\rho(\hat{y}|M,D)$), they could be both continuous, discrete (with order),  categorical variables (no well defined order, e.g. strings, pictures,...), or a mix.} Equation (\ref{genagree}) is the general form of the Bayesian Validation Metric (BVM). Because $A$ is discrete, the BVM is a probability rather than a probability density and it therefore falls in the range $0\leq p(A|M,D)\leq 1$. Equation (\ref{genagree2}) explicitly assumes that the uncertainty in the data is independent of the model, i.e. $\rho(z|M,\hat{z},D)=\rho(z|D)$, that the data $D$ does not take $\hat{z}$ or the model $M$ (that it is currently being compared to) as inputs.\footnote{In a controls system this may not be the case, as the model may interact with the system of interest. In such a case this constraint may be lifted and one should use (\ref{genagree}) instead. The joint probability $\rho(\hat{z},z|M,D)$ can be used to account for the correlations between the model (the controller or reference) and the data (the measured response of the system being controlled) in a controls setting in principle.} This is a relatively common scenario so it is stated explicitly.  The BVM may be given a geometric interpretation as the projection of two probability vectors (potentially of unequal length) in a space whose overlap is defined by the agreement kernel -- this is an inner product if $B(\hat{z},z)$ is symmetric in its arguments. The BVM may be computed using any of the well known computational integration methods.
	

	\subsection{An identical representation}
	In some cases, it is useful to work directly with the probability density $\rho(f|M,D)$, which quantifies the probability the comparison value function $f(\hat{z},z)$ takes the value $f$ due to uncertainty in its inputs. This pdf is independent of any user defined accuracy requirement. We will call this pdf the comparison value probability density, which is equal to,
	\begin{eqnarray}
	\rho(f|M,D)=\int_{\hat{z},z} \delta(f-f(\hat{z},z))\cdot\rho(\hat{z},z|M,D)\,d\hat{z}\,dz.\label{metricprob}
	\end{eqnarray}
	This is the net uncertainty propagated through the comparison value function $f(\hat{z},z)$ from the uncertain model and data comparison values. All of the expectation values that are associated with $f$ may be generated from this pdf.	
	
	If one imposes an accuracy requirement with a Boolean expression $B=B(f)$ (i.e. defining agreement according to the value of $f$), the resulting accumulated probability is the BVM. That is, the BVM, i.e. equation (\ref{genagree}), may equally be expressed as,
	\begin{eqnarray}
	p(A|M,D)=\int_f \rho(f|M,D)\cdot\Theta(B(f))\,df,\label{20}
	\end{eqnarray}
	which is proven through substitution and marginalization over $f$,
	\begin{eqnarray}
	p(A|M,D)&=&\int_f \Big(\int_{\hat{z},z} \delta(f-f(\hat{z},z))\cdot\rho(\hat{z},z|M,D)\,d\hat{z}\,dz\Big)\cdot \Theta(B(f))\,df\nonumber\\
	&=&\int_{\hat{z},z} \Theta(B(\hat{z},z))\cdot\rho(\hat{z},z|M,D)\,d\hat{z}\,dz. 
	\end{eqnarray}
		\subsection{Importance}

	The BVM allows the user to, in principle, quantify the probability the model and the data agree with one another under arbitrary comparison value functions and with arbitrary definitions of agreement. The BVM can therefore be used to fully quantify the probability of agreement between arbitrary model and data types using novel or existing comparison value functions and definitions of agreement. Thus, the problem of model-data validation may be reduced to the problem of finding the four BVM inputs in any model validation scenario.
	
			\subsection{Statistical Responsibility\label{section24}}

When using the BVM framework, one should practice statistical responsibility by explicitly stating the definition of agreement that is implemented in the validation procedure. Although the flexibility of the BVM framework is a feature, different validation metrics often have different amounts of tolerance as what constitutes ``agreement". Agreement according to one metric does \emph{not} in general imply agreement according to another. Overly tolerant definitions of agreement have little resolution power and can only be used responsibly if a large degree of non-exactness between the model and data is permissible. In principle, the definition of agreement should be just as strict or stricter than it needs to be. By explicitly stating the definition of agreement along side the BVM value, $p(A|M,D)\equiv p(A|M,D,B)$, one avoids statistical misrepresentation by not hiding one's definition of agreement. 
	

	\section{Meeting the desirable validation criterion}
	
	
	First we will describe how the BVM, equations (\ref{genagree}) and (\ref{20}), precisely match our validation criterion (1.). 
	As can be seen by equation (\ref{metricprob}), incorporated into the BVM is a statistically quantified quantitative measure that compares data and model outputs, $\rho(f|M,D)$. However, this pdf is in some sense lacking a context pertaining to the model-data pair. Not until an accept/reject rule is imparted on $\rho(f|M,D)$ does one define what is meant by \emph{agreement} in the model-data context. Thus, the BVM only becomes the probability of \emph{agreement} between the data and the model when the agreement function is also incorporated. The four BVM inputs are therefore adequate to satisfy (1.) as the BVM is a ``statistically quantified quantitative measure ($f$) of agreement $p(A|M,D)$ between model predictions and data pairs $(\hat{z},z)$, in the presence or absence of uncertainty $\rho(\hat{z},z|M,D)$".
    
    There are a few more BVM concepts worth discussing before moving forward. We will show that the BVM is capable of handling general multidimensional model-data comparisons and that there are no conceptual issues when agreement is exact, i.e. $B$ is true iff $\hat{z}=z$, in the certain and uncertain cases. We will then make comments on the sense in which the BVM adheres to the full set of six desirable validation criteria given in \cite{Liu2} by discussing the criteria that are underrepresented in (1.). 
    
	
	\subsection{Compound Booleans}
	Because Boolean operations between Boolean functions results in a Boolean function itself, the BVM is capable of handling multidimensional model-data comparisons. We will call a Boolean function with this property a ``compound Boolean". A compound Boolean function results from \emph{and}, $\wedge$, conjunctions and \emph{or}, $\vee$, disjunctions between a set of Boolean functions, e.g.,
	\begin{eqnarray}
	B(\{B_i\})=\Big(B_1(\hat{z},z) \vee B_2(\hat{z},z)\Big) \wedge \Big(B_3(\hat{z},z)\vee B_4(\hat{z},z)\Big)...,\label{Compoundbool}
	\end{eqnarray}
	where each $B_i(\hat{z},z)=B_i(f_i(\hat{z},z))$ may use a different comparison function $f_i(\hat{z},z)$. Compound Booleans using conjunctions quantifying the validity of entire model functions (random fields and/or multidimensional vectors) by assessing agreement between each of the model-data comparison field points simultaneously, i.e over the comparison points 1 \emph{and} points 2 \emph{and} so on. The compound Booleans may be factored into their constituting Boolean functions using the standard product and sum rules of probability theory after being mapped to probabilities with the agreement kernel. One should be weary when defining an \emph{and} Boolean that, if one of the Booleans is false, then the entire Boolean is false. If this strict ``all or nothing" validation requirement is not needed then other more flexible definitions of agreement may be instantiated instead (see ``BVM Examples").

	\subsection{The BVM under the conditions of exact agreement}
	We can calculate the BVM under the conditions of exact agreement in the completely certain and uncertain cases. Because the BVM is a probability rather than a probability density, the agreement kernel falls in the range $[0,1]$. Under the conditions of exact agreement, that $B$ is only true when $\hat{z}=z$,  the agreement kernel is $\Theta(B)=\Theta(\hat{z}=z)=\delta_{\hat{z},z}$, which is the Kronecker delta, i.e. it is $0$ or $1$, but has continuous labels. As it is uncommon to deal with Kronecker delta's having continuous labels under integration, we will show that the BVM gives reasonable results under the condition of exact agreement in the complete certainty as well as in the general uncertain case. 
	\paragraph{Complete certainty and exact agreement}
	Complete certainty is represented using Dirac delta pdf functions over the model and data comparison values. This gives the BVM,
	\begin{eqnarray}
	p(A|M,D)&=&\int_{\hat{z},z} \rho(\hat{z}|M,D)\Theta\Big(\hat{z}=z\Big) \,\rho(z|D) \,d\hat{z}\,dz\nonumber\\
	&=&\int_{\hat{z},z} \delta(\hat{z}-\hat{z}')\cdot\delta_{\hat{z},z}\cdot \delta(z-z')\,d\hat{z}\,dz,\label{deltacert-1}
	\end{eqnarray}
	where we are considering the model-data pair to agree iff the comparison values are exactly equal. Using the sifting property of the Dirac delta function, we find the reasonable result that,
	\begin{eqnarray}
	p(A|M,D)=\delta_{\hat{z}',z'},\label{deltacert}
	\end{eqnarray}
	which is equal to unity iff $\hat{z}'$ and $z'$, the definite values of $\hat{z}$ and $z$, are equal. 
	\paragraph{Uncertainty and exact agreement}
	In the uncertain case under the condition of exact agreement, the BVM is
	\begin{eqnarray}
	p(A|M,D)=\int_{\hat{z},z} \rho(\hat{z}|M,D)\cdot\delta_{\hat{z},z}\cdot \rho(z|D)\,  d\hat{z}\,dz.
	%
	\end{eqnarray}
	We will do the following trick to correctly interpret this integral. We will first let $B(\epsilon)$ be true if $z-\epsilon\leq \hat{z}\leq z+\epsilon$ and then take the limit as $\epsilon\rightarrow 0^{+}$ such that $\lim_{\epsilon\rightarrow0^+}B(\epsilon)\rightarrow B$ when appropriate. With this Boolean expression, the BVM is,
	\begin{eqnarray}
	p(A|M,D,\epsilon)&=&\int_{\hat{z},z} \rho(\hat{z}|M,D)\Theta\Big(z-\epsilon\leq \hat{z}\leq z+\epsilon\Big) \,\rho(z|D) \,d\hat{z}\,dz\nonumber\\
	&=&\int_{z}\rho(z|D)\Big(\int_{z-\epsilon}^{z+\epsilon} \rho(\hat{z}|M,D) \,d\hat{z}\Big)\,dz.\label{trick}
	\end{eqnarray}
	In the limit $\epsilon\rightarrow0^+$, the term in the parenthesis $\int_{z-\epsilon}^{z+\epsilon} \rho(\hat{z}|M,D) d\hat{z}\rightarrow p(\hat{z}=z|M,D)=\rho(\hat{z}=z|M,D) d\hat{z}$ by the definition of probabilities. This gives,
	\begin{eqnarray}
	p(A|M,D)&=&\int_{z}\rho(z|D)\Big( p(\hat{z}=z|M,D)\Big)dz=\Big(\int_{z} \rho(\hat{z}=z|M,D)\rho(z|D)dz\Big)\,d\hat{z}\nonumber\\
	&\equiv&\rho(\hat{z}\equiv z|M,D)d\hat{z}=p(\hat{z}\equiv z|M,D),\label{exactagree}
	\end{eqnarray}
	which is understood to be the sum of the model and the data probabilities that jointly output exactly the same values.
	We see that the BVM in this case is proportional to $d\hat{z}$, $p(A|M,D)\rightarrow \rho(\hat{z}\equiv z|M,D)d\hat{z}$, in the general case of exact agreement, and therefore the BVM goes to zero unless the pdf $\rho(\hat{z}\equiv z|M,D)\propto\delta(...)$. Thus, we recover the standard logical result for probability densities $p(x)=\rho(x)dx\rightarrow0$ unless it is offset by $\rho(x)\propto\delta(x)$. This result is easily generalized to the dependent case using $\rho(\hat{z},
	z|M,D)=\rho(z|M,D)\rho(\hat{z}|z,M,D)$. The result, equation (\ref{exactagree}), is no more surprising than (\ref{deltacert}) in principle.  
	
	Due to the vast number of possibilities for continuous valued variables, having a pathological definition of exact agreement between continuous variables does not occur in practice. In a computational setting, $dx\rightarrow \bigtriangleup x$ becomes a finite difference and these infinitely improbable agreement conceptual issues are avoided. The Bayesian model testing framework avoids these issues by evaluating posterior odds ratios, in which case the measures, $d\hat{z}$, drop out.
	
\subsection{Meeting underrepresented validation criteria}
In this subsection we will discuss how the BVM also meets the validation criteria found in \cite{Liu2}. This is done by using the derived general and special cases of the BVM for each of the criteria which are underrepresented in (1.).
    
    Perhaps the primary underrepresented criterion from \cite{Liu2} is their second. It states that ``..the criteria used for determining whether a model is acceptable or not should not be a part of the metric which is expected to provide a quantitative measurement only." We argue that the functional form of the BVM presented in equation (\ref{20}) clearly demonstrates this feature as it factors into $\rho(f|M,D)$ and $\Theta(B(f))$. The comparison function $f(\hat{z},z)$ represents the ``objective quantitative measure" from their first criterion that is separate from the accept/reject rule, which is our agreement function $B(f)$ -- both of which require definition to ultimately evaluate the validity of a model. We see it as advantageous to quantify the probability the model is accepted or rejected through $B(f)$ due to the uncertainty in the value of $f$, which is the general case, and which gives the BVM as the result.   As all of the validation metrics presented in \cite{Liu2} (and more) will be shown to be representable with the BVM, and thus placed on the same footing, we find our language of ``comparison function" and ``agreement function" to ultimately be more useful than a language that only considers comparison functions (without accept/reject rules) to be the validation metrics.
    
    The third criteria in \cite{Liu2} is that ideally the metric should ``degenerate to the value from a deterministic comparison between scalar values when uncertainty is absent". This is indeed the case as can be seen in equation (\ref{agree}) or in equations (\ref{genagree}) and (\ref{20}) by utilizing Dirac delta pdfs similar to their application in equation (\ref{deltacert-1}).

The fifth desirable validation criteria in \cite{Liu2} states that artificially widening probability distributions should not lead to higher rates of validation. They find all but the frequentist metric to have this undesired feature; however, we later see that the frequentist metric may be considered a special case of the reliability metric (when reasonable accuracy requirements are imposed), meaning artificial widening can lead to higher rates of validation for more general instances of the frequentist metric. 

Further, we argue that artificially introducing uncertainty for the express purpose of passing a validation test is indistinguishable from scientific misconduct. If there is objective reason to include more uncertainty into the analysis or if the circumstance for what constitutes validation has changed due to a change of context -- and it happens to improve the rate of validation -- so be it. This is a different context, model, or state of uncertainty than was originally proposed so different rates of acceptance should be expected. Reducing the uncertainty of either the data or the model (the inputs) through additional measurements or changing the model may later prove the model valid or invalid when it may have been initially accepted. Thus, to meet this validation criteria, we simply assume the user is not engaging in scientific misconduct.
    
Finally, due to the results of the ``Compound Booleans" section, their sixth criterion is met. Because the BVM (\ref{genagree}) can be used to assess single or multidimensional controllable settings (see footnote number 3.) we can perform global function validity (in or out of a controls setting). As they note, ``This last feature is critical from the viewpoint of engineering design". 

Thus, the BVM satisfies both our validation criterion and the six desirable validation criteria outlined in \cite{Liu2}. This was accomplished by representing model-data validation as an inference problem using the four BVM inputs.

    	

		
	\section{Representing and generalizing the known validation metrics with the BVM}
	This section is a review of the material found in Appendix A. The following validation metrics will be represented with the BVM, which are then are improved, generalized, and/or commented on: reliability/probability of agreement, improved reliability metric, frequentist, area metric, pdf comparison metrics, statistical hypothesis testing,  and Bayesian model testing.
	\subsection{Representating the known validation metrics with the BVM}
	Table 1 outlines the values of the four BVM inputs that result in the BVM representing each of the well known validation metrics as special cases. The following notation is used for the comparison values $(\hat{z},z)$. The brackets $\expt{...}$ denote expectation values, $\mu$'s denote averaged values, $\hat{y},y$ denote single values, $\hat{Y},Y$ denote multidimensional (or many valued) values, $F_y,F_{y}$ denote cumulative distribution functions ($F_y=\int_{-\infty}^{\hat{y}}\rho(\hat{y}|M,D)\,dy$), $S_{\hat{y}},S_{y}$ denote test statistics, and $[-c_{\alpha},c_{\alpha}]$ denotes the $1-\alpha$ confidence interval of the data. In the agreement function column, an element listed as $B(f)$ means the creators of the metric intentionally left the definition of agreement unspecified; however, it is natural to assume it is a function of the comparison function $f$. 

		Table 1 shows the specification of the four BVM inputs that give the other validation metrics as special cases. It also summarizes some of the similarities and difference between the known validation metrics. In particular, by looking at the validation metrics with the same type of comparison values, i.e. the reliability and frequentist or the improved reliability and Bayesian model testing, we can compare them directly. We see that if one lets the frequentist metric allow for more general input probability distributions and the use of a reasonable agreement function (i.e., $B(f)$ is true if $|f|<\epsilon$), then the frequentist metric is the reliability metric. Further, in Bayesian model testing, if the agreement function $B(f)$ is loosened to accept $f<\epsilon$, than the pdf's that appear in the Bayesian model testing framework are equal to the improved reliability metric. This information improves the objectivity of the current validation procedure because we now have a map between validation metrics that were originally thought to be different.

\begin{table}[H]
\begin{center}

    \caption {Specification of the four BVM inputs that give the other validation metrics as special cases.}

\begin{tabular}{ |c|cc|c|c|c|c| }
\hline
&\multicolumn{2}{ |c| }{Comp. Values}&\multicolumn{2}{ |c| }{Probs.}&\multicolumn{1}{ |c| }{Comp. Func.}&\multicolumn{1}{ |c| }{Agree. Func.} \\\hline\hline
BVM& $\hat{z}$& $z$ & $\rho(\hat{z}|M,D)$& $\rho(z|D)$&\multicolumn{1}{ |c| }{$f(\hat{z},z)$}&\multicolumn{1}{ |c| }{$B(f)$} \\
\hline\hline
Reliability& $\expt{\hat{y}}$& $\mu_{y}$ & $\rho(\expt{\hat{y}}|M,D)$& $\rho(\mu_{y}|D)$&\multicolumn{1}{ |c| }{$|\expt{\hat{y}}-\mu_{y}|$}&\multicolumn{1}{ |c| }{$f<\epsilon$} \\
\hline
Imp. Reli.& $\hat{Y}$& $Y$ & $\rho(\hat{Y}|M,D)$& $\rho(Y|D)$&\multicolumn{1}{ |c| }{$|\hat{Y}-Y|$}&\multicolumn{1}{ |c| }{$f<\epsilon$} \\
\hline
Frequentist & $\expt{\hat{y}}$& $\mu_{y}$ & $\delta(\expt{\hat{y}}-\expt{\hat{y}}')$& Stud. t&\multicolumn{1}{ |c| }{$\expt{\hat{y}}-\mu_{y}$}&\multicolumn{1}{ |c| }{$B(f)$} \\
\hline
Area & $F_y$& $F_{y}$ & $\delta(F_y-F_{\hat{y}}')$& $\delta(F_{y}-F_{y}')$&\multicolumn{1}{ |c| }{$\int_y|F_y-F_{y=\hat{y}}|$\scriptsize{dy}}&\multicolumn{1}{ |c| }{$B(f)$} \\
\hline
Pdf Comp. & $\rho_M$& $\rho_D$ & $\delta(\rho_M-\rho_M')$& $\delta(\rho_D-\rho_D')$&\multicolumn{1}{ |c| }{$G(\rho_D||\rho_M)$}&\multicolumn{1}{ |c| }{$B(f)$} \\
\hline
Stat. Hyp. & $S_{\hat{y}}$& $S_{y}$ & $\rho(S_{\hat{y}}|M=D)$&$\rho(S_{y}|D)$&\multicolumn{1}{ |c| }{$S_{\hat{y}}$}& \multicolumn{1}{ |c| }{$S_{\hat{y}}\in [-c_{\alpha},c_{\alpha}]$} \\
\hline
Bayes Model & $\hat{Y}$& $Y$ & $\rho(\hat{Y}|M,D)$& $\rho(Y|D)$&\multicolumn{1}{ |c| }{$|\hat{Y}-Y|$}&\multicolumn{1}{ |c| }{$f=0$} \\
\hline
\end{tabular}
    \caption*{\\The column headings are the four BVM input values: Comparison Values $(\hat{\hat{z}},z)$, Probabilities $(\rho(\hat{z}|M,D),\rho(z|D))$, Comparison Function $f=f(\hat{z},z)$, and the Boolean Agreement Function $B(f)$. The row headings read: Reliability, Improved Reliability, Frequentist, Area, Pdf Comparison Metrics, Statistical Hypothesis Test, and Bayesian Model Test. The denoted data probability for the average in the frequentist metric, Stud. t, is the Student t distribution.}
\end{center}
\end{table}

\begin{table}[H]
\begin{center}
    \caption {BVM representation of the special cases using the $f$'s specified in Table 1.}
\begin{tabular}{ |c|c| }
\hline
& BVM \\
\hline\hline
BVM& $\int_f \rho(f|M,D)\cdot\Theta(B(f))\,df$\\\hline\hline
Reliability& $\int_{f} \rho(f|M,D)\cdot\Theta(f<\epsilon)\,df=r $\\
\hline
Imp. Reli.& $\int_{f} \rho(f|M,D)\cdot\Theta(f<\epsilon)\,df=r_i$\\
\hline
Frequentist &  $\int_{\mu_{y}} \rho(\mu_{y}|D)\cdot\Theta(B(f(\expt{\hat{y}}',\mu_{y})))\,d\mu_{y}$\\
\hline
Area & $\Theta(B(f(F_y',F_{y}'))$ \\
\hline
Pdf Comp. & $\Theta(B(f(\rho_M',\rho_D'))$ \\
\hline
Stat. Hyp. & $\int_{S_{\hat{y}}} \rho(S_{\hat{y}}|M=D)\cdot\Theta(S_{\hat{y}}\in [-c_{\alpha},c_{\alpha}])\,d S_{\hat{y}}=1-\alpha$ \\
\hline
Bayes Model &$\int_{f} \rho(f|M,D)\cdot\Theta(f=0)\,df=p(\hat{Y}\equiv Y|M,D)$ \\
\hline
\end{tabular}
\end{center}
	\end{table}
	
Table 2 shows the resulting BVM using the specifications listed in Table 1. The value $r$ is the standard notation for the reliability metric \cite{Rebba} and we use $r_i$ for the improved reliability metric \cite{Sankararaman2}. The BVM represents each of the known validation metrics as a probability of agreement between the model and the data from equation (\ref{genagree}). As no agreement function is specified directly for the frequentist and area metric, the problem is under constrained so the agreement functions are left as general functions over the comparison function $B(f)$. Thus, for any chosen agreement function, the BVM quantifies their probability of agreement. The remaining metrics all do specify (or indicate) an agreement function, and thus, have specified all of the information required to compute the BVM.

The statistical hypothesis test is perhaps a bit out of place among the validation metrics. First, note that the comparison function for statistical hypothesis testing is not a function of both the data and the model. Further note, the model pdf used for statistical hypothesis testing assumes the null hypothesis is true, which in our language is the assumption that $\rho(S_y|M,D)=\rho(S_y|M=D)$, i.e. that the pdf of the model is equal to the pdf of the data. This shows how statistical hypothesis is a bit out of place here among the validation metrics because here we are attempting to validate a model, usually with its own quantified pdf, rather than, perhaps irresponsibly, assuming it is equal the data pdf before validating that to be the case. This causes standard statistical hypothesis pitfalls, such as type I (rejecting the null hypothesis when it is true) and type II errors (failing to reject the null hypothesis when it is false), to be carried over into BVM, which is unwanted. Several comments are made in Appendix A.5 on this issue.

A perhaps surprising result is the proposed functional form of the BVM that represents Bayesian model testing $p(A|M,D)=p(\hat{Y}\equiv Y|M,D)$, which is the Bayesian evidence. This is the probability the uncertain model and data happen to output exactly the same values. Usually what is discussed when reviewing Bayesian model testing is the Bayes posterior odds ratio, i.e. the ``Bayes Ratio", $$R=\frac{p(M|Y)}{p(M'|Y)}\propto\frac{p(Y|M)}{p(Y|M')},$$ which tests one model $M$, i.e. for validation, against another model $M'$. However, in validation metric problems, we are first interested in considering the validation of a single model -- the ratio is an extra bit of inference. In Appendix A.6 we show that the BVM result of $p(\hat{Y}\equiv Y|M,D)$ is exactly what we mean by $p(Y|M)$ in the numerator of the Bayes factor,\footnote{It should be noted that our notation for $D$ differs from the notation typically used in Bayesian model testing. Their $D$ is equal to our data $Y$, while our $D$ refers to context ``as having come from the data or experiment rather than the model". } which effectively quantifies the validation of a single model against data $Y$, all quantified under uncertainty.

		





	\subsection{Generalizations and Improvements to the known validation metrics with the BVM}
	
	The BVM offers several avenues to either generalize or improve many of the metrics. The types of generalizations the BVM offer pertain to generalizing the comparison values, comparison functions, definitions of agreement, and/or generalizing deterministic comparison values and metrics to the uncertain case. These generalizations are only useful if quantitative statements can be made on their behalf -- in such a case, these generalizations are improvements. We will give a brief review of the improvements we found below, but the full discussion is located in Appendix A. By making generalizations or improvements to each of the known validation metrics as implied by the BVM, each metric can be made to satisfy our validation criterion as well as the six desirable validation criteria in \cite{Liu2}, due to the results of Section 3.
	
 Appendix A.1 uses the BVM to show that the reliability metric and the improved reliability metric can be generalized to compare values without a unique order, such as strings, in principle. This involves creating an agreement function over sets of values (such as synonymous sets of strings), rather than in a continuous interval, that may be considered to ``agree". 
 
 Appendix A.2 derives the frequentist validation metric and generalizes it to the case where both the model and data expectation values are uncertain. The frequentist metric assumes the model outputs are known with certainty, which may or may not be true. If a model is stochastic, the model pdfs may be estimated with Monte Carlo or other uncertainty propagation methods that quantify the pdf directly.
 
 Appendix A.3 shows that the area metric may be cast as a special case of the BVM. The area metric involves quantifying the difference between model and data cumulative distributions on a point to point basis; thus, the comparison values $(\hat{z},z)$ are cumulative distributions themselves. The comparison values are assumed to be known with complete certainty, which in the case of cumulative distributions of data is often difficult to argue. Any quantifiable uncertainty in the cumulative distributions may integrated over, which generalizes the area metric to situations when the model and/or the data cumulative distributions are uncertain. A drawback is that the BVM in these cases may be very computationally intensive and would likely need to be approximated using a random sampling or discretization scheme. A binned pdf metric is put forward to potentially reduce the computational complexity toward quantifying this generalized area validation metric. This applies similarly to the pdf comparison metrics in Appendix A.4.
 
  In Appendix A.5 we invent an improved statistical hypothesis test using the BVM that takes into account both model and data pdfs called the ``statistical power BVM". Because in principle we have a model output pdf in model validation problems $\rho(\hat{y}|M,D)$, we can use it (in place of assuming the null hypothesis is true) to remove the possibility of both type I and type II errors. 
  
  In the statistical power BVM, the model and the data are defined to agree if their test statistics both lie within one another's confidence intervals (or ``confidence sets" as explained in Appendix A.5).  The statistical power BVM becomes the product of the statistical powers of the model and data, denoted $p(A|M,D)=(1-\beta_D(\hat{\alpha}))\cdot(1-\beta_M(\alpha))$ in equation (\ref{improvedSHT}). Further comments are made about how systematic error (defined as when a test statistic lies outside of its \emph{own} confidence interval) may be removed. 
  
  It is concluded that the statistical power BVM has a relatively low resolving power compared to other BVMs. This is because large confidence intervals imply large tolerance intervals for acceptance. For this reason, statistical hypothesis testing should only be used for validation in situations when a high degree of nonexactness between model and data test statistics is permissible and the pdf's have very thin tails. This BVM does, however, have a greater resolution than the classical hypothesis test as was proved in the appendix and will be demonstrated next section.
  
  Appendix A.6 finds that Bayesian model testing has the highest possible resolving power because the model and the data are defined to agree only if their values are exactly equal. This is the reverse of what was concluded about statistical hypothesis testing.  
  
	
	Further in Appendix A.6, we argue that, analogous to the Bayesian model testing framework, nothing prevents us from constructing what we call the BVM factor.
	The BVM factor is,
	\begin{eqnarray}
	K(B)=\frac{p(A|M,D,B)}{p(A|M',D,B)},
	\end{eqnarray}
	which is a ratio of the BVMs of two models under arbitrary definitions of agreement $B$. Using Bayes Theorem, $p(M|A,D,B)=\frac{p(A|M,D,B)P(M|D,B)}{P(A|D,B)}$, we may further construct the BVM ratio,
	\begin{eqnarray}
	R(B)=\frac{p(M|A,D,B)}{p(M'|A,D,B)}=\frac{p(A|M,D,B)P(M|D,B)}{p(A|M',D,B)P(M'|D,B)}=K(B)\frac{P(M|D,B)}{P(M'|D,B)},\label{BVMratio}
	\end{eqnarray}
	for the purpose of comparative model selection under a general definition of agreement $B$.
	The ratio $\frac{P(M|D,B)}{P(M'|D,B)}$ is the ratio prior probabilities of $M$ and $M'$, which analagous to Bayesian model testing, if there is no reason to suspect that one model is a priori more probable than another, one may let $\frac{P(M|D,B)}{P(M'|D,B)}=1$, and then $R(B)\rightarrow K(B)$ in value. 
	
	Thus, using the BVM ratio, we can perform \emph{general model validation testing under arbitrary definitions of agreement and with any reasonable set of comparison functions}. The BVM ratio therefore generalizes the Bayesian model testing framework. This will be utilized in next section. 
	
	Finally we wanted to add a note about how one may mitigate the sharpness of the indicator function without using fuzzy logic while also allowing close models to be somewhat accepted. As we have seen, it is natural to use a threshold Boolean parameter $\epsilon$ to help define the boundary of agreement through $B(f\leq\epsilon)$. Such a BVM takes the form,
	\begin{eqnarray}
	p(A|M,D,\epsilon)=\int_f \rho(f|M,D)\cdot\Theta(B(f\leq\epsilon))\,df,
	\end{eqnarray}
	where $\Theta\Big(...\Big)$ instantaneously drops to zero for $f>\epsilon$. One may soften the boundary by allowing $\epsilon$ itself be an uncertain quantity, which means one allows their definition of agreement to be somewhat uncertain (which often times can be reasonably claimed). As an example, let this uncertainty be $
\rho(\epsilon)=\lambda\exp(-\lambda (\epsilon-\epsilon'))$ for $\epsilon'>\epsilon$ and zero otherwise, where $\lambda$ is positive. Marginalizing over this BVM then gives,
		\begin{eqnarray}
	p(A|M,D)&=&\int_{\epsilon}p(A|M,D,\epsilon)\rho(\epsilon)\,d\epsilon\nonumber\\
	&=&\int_f \rho(f|M,D)\cdot\Big[\Theta\Big(B(f\leq\epsilon')\Big)+\Theta\Big(B(f>\epsilon')\Big)e^{-\lambda(f-\epsilon')}\Big]\,df,\label{marginalizeme}
	\end{eqnarray}
	which allows some $f$'s to be accepted outside the agreement region defined by $f\leq\epsilon'$, but with a exponentially decaying probability. Other potentially useful $\epsilon$ pdfs include, but are not limited to: negative slope linear, Gaussian, or decaying sigmoidal. None of these $\epsilon$ type distributions were needed to obtain the results of the previous sections explicitly; however, these types of assumptions may have been part of the decision process made implicitly by a practitioner while performing model validation. 

\section{BVM Examples}

In this section we invent and quantify three novel validation metrics using the BVM to highlight the conceptual clarity, flexibility, and capacity of our framework.  
\subsection{The Statistical Power BVM}
Here we consider the statistical power BVM proposed in \ref{stathypsection} and reviewed in the previous section. This metric defines agreement as occurring when both the model and data comparison values are within one another's confidence intervals, simultaneously. The BVM for this metric is the product of the statistical powers of the model and the data $p(A|M,D)=(1-\beta_D(\hat{\alpha}))\cdot(1-\beta_M(\alpha))$, which is calculated in equation (\ref{improvedSHT}). We contrast this with the standard statistical hypothesis test that, after assuming the model is correct, $M=D$, finds the probability the model lies within the data's confidence interval equal to $1-\alpha$. In the statistical hypothesis testing, one then proceeds to check the actual model output and speculates about type I and II errors. As discussed in the Appendix, we do not assume $M=D$ before validation and therefore type I and II errors are avoided. Rather, we let the statistical power BVM decide whether or not the model is valid. This provides a more informative validation procedure.

\begin{figure} [H]
 
  \includegraphics[scale=.55]{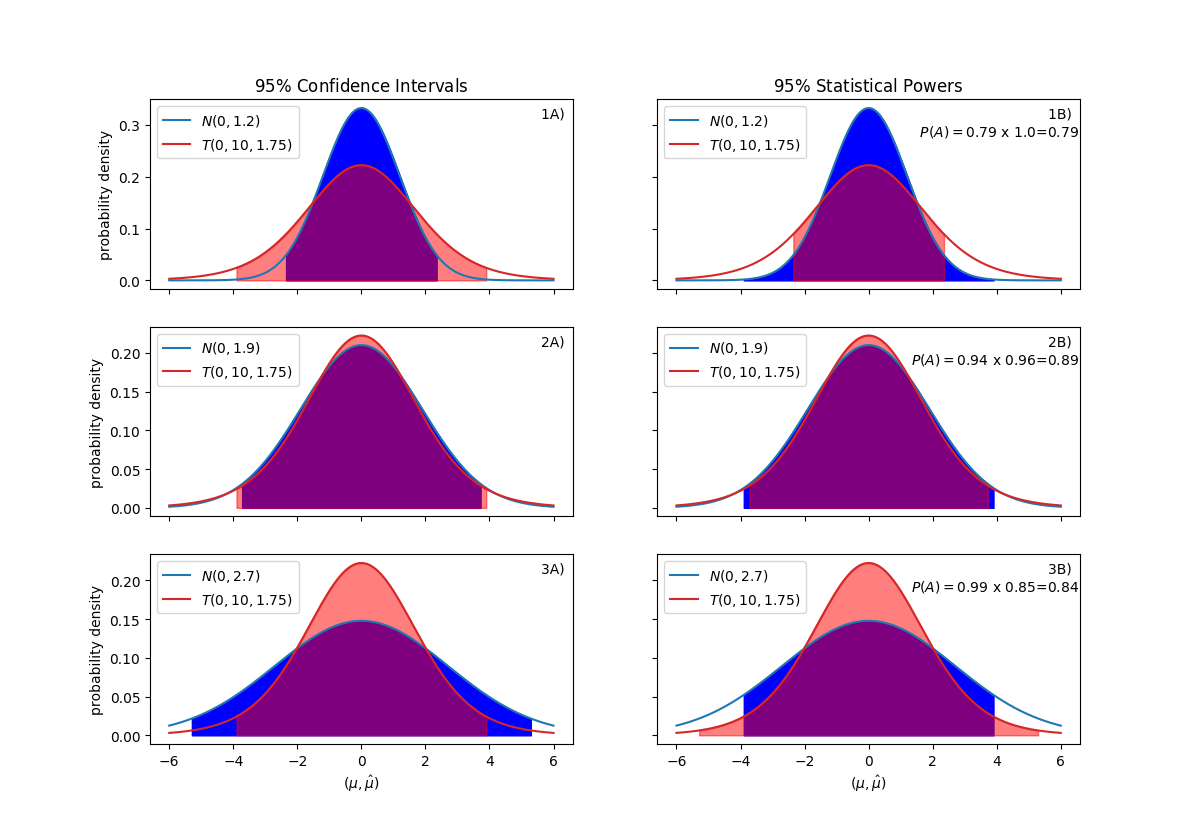}

\caption{\label{bvmstatpow} The shaded regions in Column A depict the $95\%$ confidence interval of each distribution, respectively. Because the data distribution is the same in each figure and because the statistical hypothesis test is independent of the proposed model due to assuming the hypothesis $M=D$, each model is equally valid by that test when it is clear that the model in row B is preferable. The shaded regions in Column B depict the statistical power of the distributions -- the $95\%$ confidence intervals from each distribution is shaded (integrated) in the other's pdf. The statistical power BVM (denoted $P(A)$ in column B) is calculated for each model and indeed the model in row 2 is found to be preferable as it has the highest probability of agreement. }
\end{figure}

Figure \ref{bvmstatpow} depicts a typical statistical hypothesis test scenario that is designed to check the validity of an uncertain model average prediction $\hat{\mu}$ (in blue) against an uncertain data average prediction $\mu$ (in red). The data's $\mu$ is $t$-distributed the same in each subfigure according to $T(\overline{y},n-1,\overline{s})=T(0,10,1.75)$ where $(\overline{y},n,\overline{s})$ are the sample mean, number of collected data points, and the sample standard deviation, respectively. Each row depicts a normal distribution model centered at $0$, but with increasing model variance per row.

\subsection{The $(\expt{\epsilon},\beta_D)$ BVM}
	

We invent a novel compound Boolean that defines agreement as when the model passes an average square error threshold of $\langle\epsilon\rangle$ \emph{and} a check for probabilistic model representation. The later is imposed by requiring that $95\%\pm 4\%$ of the uncertain data lies inside the model's $1-\hat{\alpha}=95\%$ confidence interval, i.e. $1-\beta_D(\hat{\alpha})\sim 95\%$. The $\pm 4\%$ tolerance was chosen such that overly uncertain models would be marked as ``not agreeing" as they would be able to guarantee that $100\%$ of the data lie within their overly wide confidence intervals. We call this compound Boolean the $(\langle\epsilon\rangle,\beta_D)$ Boolean. The BVM in this case is
	\begin{eqnarray}
	p(A|M,D,\langle\epsilon\rangle,\beta_D)=\int_{\hat{Y},Y} \rho(\hat{Y}|M,D)\cdot\Theta\Big(B(\hat{Y},Y,\langle\epsilon\rangle,\beta_D)\Big) \cdot\rho(Y|D) \,d\hat{Y}\,dY, 
	\end{eqnarray}
where the compound Boolean $B(\hat{Y},Y,\langle\epsilon\rangle,\beta_D)$ is equal to,
	\begin{eqnarray}
B\Big(\frac{1}{N}\sum_i|y_i-\hat{y}_i|\leq\langle\epsilon\rangle\Big)\wedge B\Big(0.91\leq \frac{1}{N}\sum_i\Theta(y_i\in [-c_{\hat{\alpha}},c_{\hat{\alpha}}]_i)\leq 0.99\Big),
	\end{eqnarray}
$N$ is the number of data points in $\{y_i\}=Y$, and $[-c_{\hat{\alpha}},c_{\hat{\alpha}}]_i$ is the model's $95\%$ confidence interval at comparison location $x_i$. We treat the model confidence intervals as certain quantities, which can be achieved effectively through enough Monte Carlo (MC) simulation of the model output pdfs (although this stipulation can be removed if needed). 

Although the mathematical notation for the compound Boolean is a bit cumbersome, it is relatively easy to implement programmatically using \emph{if} statements. This ease of programming allows the BVM to have a large capacity for representing complex and abstract validation scenarios in practice. 

\begin{figure} [H]
 
  \includegraphics[scale=.55]{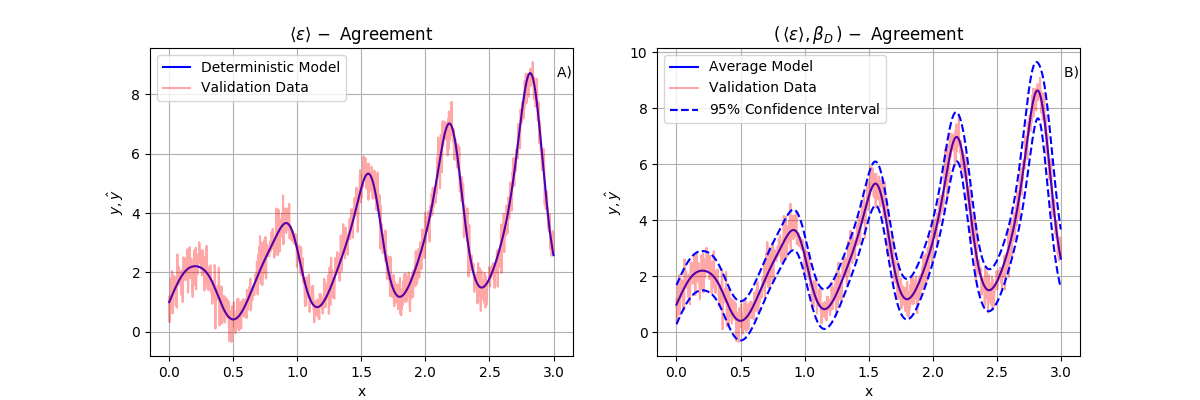}
  
\caption{\label{epsilon_beta} The deterministic model plotted in Figure \ref{epsilon_beta}A) satisfies the average error validation requirement with $P(A|\expt{\epsilon})=.99$ when a threshold of $\expt{\epsilon}=0.46$. This result is logical because the congregate standard deviation of the data is itself the combination of the aleatoric and epistemic uncertainties is $\sqrt{0.4^2+0.2^2}\approx 0.45$. This deterministic model fails to predict the uncertain fluctuations of the data because determinisic models have confidence intervals with zero width. Thus the deterministic model fails to agree according to the $(\expt{\epsilon},\beta_D)$ Boolean and one finds $P(A|\expt{\epsilon},\beta_D)=0$ for any value of $\expt{\epsilon}$. The uncertain model depicted in Figure \ref{epsilon_beta}B) is able to pass both agreement definitions; however, our choice to evaluate each probable model path against the epistemic uncertain data rather than just the average model increases the treshold to about $\expt{\epsilon}=0.9$ before an agreement probability of about $P(A|\expt{\epsilon})=.96$ is achieved. When instead evaluating $\expt{\epsilon}$ against the average model, we obtained similar results to the deterministic model for this Boolean $\expt{\epsilon}\sim0.46$ and $P(A|\expt{\epsilon})=.99$. The $(\expt{\epsilon},\beta_D)$ BVM for the uncertain model is $P(A|\expt{\epsilon},\beta_D)=0.93$, because the uncertainty in the data more or less agrees with the confidence interval provided by the model. This model can be tested for agreement against other models or model parameter distributions for their respective definitions of agreement. }
\end{figure}

Expressing the BVM as an expectation value over $\rho(Y|D)\rho(\hat{Y}|M,D)$,
	\begin{eqnarray}
	p(A|M,D,\langle\epsilon\rangle,\beta_D)=E\Bigg[\Theta\Big(B(\hat{Y},Y,\langle\epsilon\rangle,\beta_D)\Big)\Big]\sim \frac{1}{K}\sum_{k=1}^K\Theta\Big(B(\hat{Y}_k,Y_k,\langle\epsilon\rangle,\beta_D)\Big),
	\end{eqnarray}
allows one to compute the integral using standard statistical methods like MC. We use MC and $K=3000$ samples in this toy example. In Figure \ref{epsilon_beta} we implement the $(\langle\epsilon\rangle,\beta_D)$ Boolean and show that it is able to quantify both average error and a model's probabilistic representation of uncertain data, simultaneously. 

We consider data that is generated from,
\begin{eqnarray}
y(x)=1 + x\exp(-\cos(10x)) + \sin(10x)+\epsilon_a(x),
\end{eqnarray}
where $\epsilon_a(x)\sim N(0,0.4^2)$ represents aleatoric stochastic uncertainty due to the inherent randomness of the system.  An instance of the aleatoric data $Y$ without measurement uncertainty is depicted in red in Figure \ref{epsilon_beta}A) and B). We consider the data to have an additional epistemic measurement uncertainty $\epsilon_e(x)\sim N(0,0.2^2)$ that contributes to the probability of whether or not the model agrees with the plotted instance of the aleatoric data $Y$.\footnote{Our framework is not limited to investigating other validations scenarios, e.g., if one has many aleatoric data path instances $\{Y\}$ (or enough statistics about the data pdf) that one would like to validate or use for modeling testing in the congregate.}

The models plotted in Figure \ref{epsilon_beta}A) and Figure \ref{epsilon_beta}B) are generated from,
\begin{eqnarray}
\hat{y}(x;a,b,c,d,f,g)=a + bx\exp(-c\cos(dx)) + f\sin(gx),
\end{eqnarray}
where $(a,b,c,d,f,g)$ are the model parameters. In Figure \ref{epsilon_beta}A), a deterministic model is considered and plotted in blue by treating the model parameters as completely certain numbers $(a,b,c,d,f,g)=(1,1,1,10,1,10)$. In Figure \ref{epsilon_beta}B), we consider an uncertain model by treating the model parameters as uncertain values drawn from a multivariate Gaussian distribution having averages $(\mu_a,\mu_b,\mu_c,\mu_d,\mu_f,\mu_g)=(1,1,1,10,1,10)$ and standard deviations $(\sigma_a,\sigma_b,\sigma_c,\sigma_d,\sigma_f,\sigma_g)=(0.35,0.3,0.3,0.3,0.3,0.3)$. We evaluate the probability these data and model pairs pass the $\langle\epsilon\rangle$ Boolean verses the $(\langle\epsilon\rangle,\beta_D)$ Boolean by calculating their respective BVM's.

\subsection{Exploring the BVM Ratio with the $(\gamma,\epsilon)$ BVM}
In this subsection we invent an agreement function to represent the visual inspection an engineer might perform graphically and use the BVM ratio for model selection under this definition of agreement. 
By quantifying this, a practitioner could visually validate a model without actually looking at the model-data pair, which can be helpful for high dimensional spaces that are beyond human comprehension/visualizability. We will proceed by introducing this agreement function and some simple models to test it on. We will then quantify this measure using the BVM in the completely certain and uncertain cases. The main purpose of this example is to explore the BVM Ratio while showcasing the conceptual flexibility of the framework. 

To quantify something resembling the visual inspection an engineer might make graphically, we use two main criteria. We define the model to be accepted if \emph{most} of the model and data point pairs lie relatively close to one another \emph{and} if none of the point pairs deviate too far from one another. We therefore consider a compound Boolean $B(\hat{Y},Y,\gamma,\epsilon)$ that is true if a percentage larger than $\gamma$\% ($\sim$ 90\%)  of the model output points $\hat{Y}$ lie within $\epsilon$ of the data $Y$ \emph{and} 100\% of the model output points lie within some multiple $m\epsilon$ of the data, which rules out obvious model form error. We will call this compound Boolean the $(\gamma,\epsilon)$ Boolean.  The values $\gamma\%$, $\epsilon$, and $m$ can be adjusted to the needs of the modeler. It should be noted that the $\epsilon$ in this metric makes point by point evaluations as opposed to the average $\expt{\epsilon}$ Boolean used in the previous example.  We will perform the analysis for a variety of $\gamma$ and $\epsilon$ values to explore the limits of the metric. 



We will calculate the BVM for two different order polynomial models that approximate $N$ data points taken from the cosine function $y_i=\cos(x_i)$, as an illustration. The points are evenly spaced in the range $x_i\in[0,\pi]$. The first model $\hat{y}^{(1)}_i=M^{(1)}(x_i; a,b,c)=a+bx_i^2+cx_i^4$ has uncertain parameters $(a,b,c)$ and the second model $\hat{y}^{(2)}_i =M^{(2)}(x_i; a,b,c,d)=a+bx_i^2+cx_i^4 +dx_i^6$ has uncertain parameters $(a,b,c,d)$.  

To formulate the BVM, we still need to formulate the model and data probability distributions. Because the Boolean expression $B(\hat{Y},Y)$ is over the entire model and data functions, the model probability distribution is $p(\hat{Y}|M,D)$ and data probability distribution is $p(Y|D)$. These are joint probabilities over all of the points $(\hat{Y}=\{y_i\}$, $Y=\{y_i\})$ that constitute a particular path $(\hat{Y}, Y)$ of the model or data, respectively. Because both models are linear in the uncertain coefficients $(a,b,c,d)$, there is a one to one correspondence from the set of model parameters $(a,b,c,d)$ to the set of the possible paths $\hat{Y}_{a,b,c,d}$ (given $N$ is greater than the number of independent coefficients). This makes the uncertainty propagation from the uncertain model parameters to the full joint probability of the points on a path simple and results in the joint probability of paths being equal to the joint probability of the uncertain input model parameters. For simplicity we will let $p(a,b,c,d)=N(\mu_{a},\sigma_a)N(\mu_{b},\sigma_b)N(\mu_{c},\sigma_c)N(\mu_{d},\sigma_d)$, where $N(\mu,\sigma)$ is a normal distribution of average $\mu$ and standard deviation $\sigma$, such that is,

 $$p(\hat{Y}_{a,b,c,d}|M_j)=\frac{1}{Z}\exp\Big(-\frac{(a-\mu_{a})^2}{2\sigma^2_a}-\frac{(b-\mu_{b})^2}{2\sigma^2_b}-\frac{(c-\mu_{c})^2}{2\sigma^2_c}-\frac{(d-\mu_{d})^2}{2\sigma^2_d}\Big).$$ Because the problem is well understood, we discretize the integrals rather than estimating them with MC. After discretization, the $(\gamma,\epsilon)$ BVM for each model is,
\begin{eqnarray}
p(A|M_j,D,\gamma,\epsilon)=\sum_{\hat{Y},Y} p(\hat{Y}|M_j)\cdot\Theta(B(\hat{Y},Y,\gamma,\epsilon))\cdot p(Y|D).
\end{eqnarray} 
In principle $\epsilon=(\epsilon_1,..,\epsilon_N)$ is an $N$ dimensional vector where each $\epsilon_i$ may be adjusted to impose more or less stringent agreement conditions on a point to point basis, which may be used to enforce reliability in regions of interest. In our example we let all of the components of $\epsilon_i$ be equal. If the standard deviation of each data point $\sim\hat{\sigma}_i$ (aleatoric and/or measurement uncertainty) in the joint data pdf $p(Y|D)$ is much less than $\epsilon_i$ and $N$ is large, one may approximate the BVM as,
\begin{eqnarray}
p(A|M_j,D,\gamma,\epsilon)\approx\sum_{\hat{Y}} p(\hat{Y}|M_j)\cdot\Theta(B(\hat{Y},Y',\gamma,\epsilon)),
\end{eqnarray} 
which can greatly reduce the number of combinations one must calculate by effectively treating the data as known, deterministic, and equal to $Y'$. We will use this approximation as it does not take away from the main point of this example.

We will use the following numerics. In the completely certain case, we will let the parameters be the Taylor series coefficients $(a,b,c,d)=(1,-\frac{1}{2!},\frac{1}{4!},-\frac{1}{6!})$ ($d=0$ for model 1) and in the uncertain case we let each coefficient have Gaussian uncertainty centered at their Taylor series coefficients with standard deviations $(\sigma_a,\sigma_b,\sigma_c,\sigma_d)=(0.1,0.05,0.005,0.0005)$ (and where $\sigma_d=0$ for model 1). We let each model output path have $N = 50$ points and we allow for $20$ possible values per parameter $(a, b,c,d)$, which results in $20^3=8000$ possible paths for model 1 and $20^4=160,000$ for model 2. We let $\gamma$ vary between $75\%$ and $100\%$ using an increment of $1\%$ and let $\epsilon$ vary between $0$ and $1$ using an increment of $0.01$. The value of $m$ was chosen to be equal to $5$, which imposes that no model path can have points that are greater than $5\epsilon$ away while still be considered to agree with the data. 

The BVM probability of agreement values as a function of the Boolean function parameters $(\gamma,\epsilon)$ are plotted in Figure \ref{2DeterministicPlot} for model 1 and model 2 in the completely certain case:
  \begin{figure} [H]
\indent 
  \includegraphics[scale=0.42]{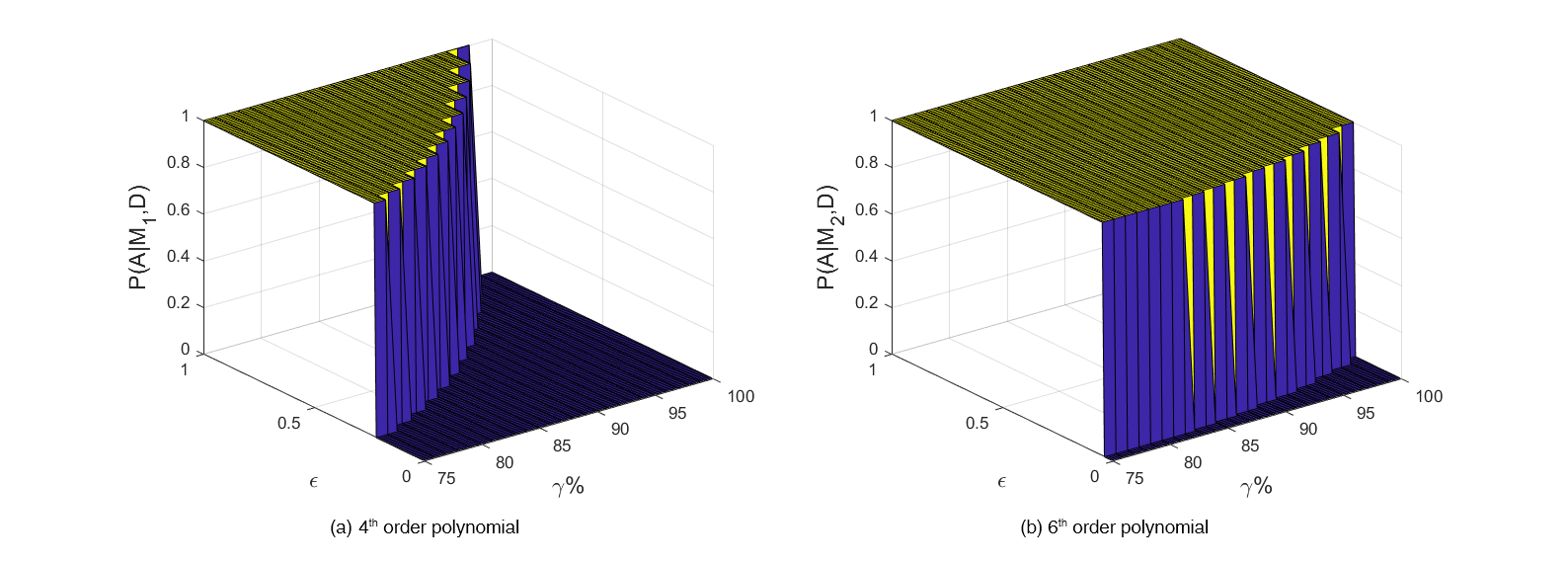}
\caption{\label{2DeterministicPlot} Completely Certain Case: The BVM probability of agreement between model 1 and 2 with the data is plotted in the space of $(\gamma,\epsilon)$. The results for model 1 ($4^{\mathtt{th}}$ order polynomial) is plotted on the left and model 2 ($6^{\mathtt{th}}$ order polynomial) is plotted on the right. Because here the models are deterministic, the BVM probability of agreements for each $(\gamma,\epsilon)$ pair is either zero or one. As expected, model 2 better fits the data in the space of $(\gamma,\epsilon)$ as it has more BVM values equal to one than model 1 as it is overall closer to the cosine function being that it is the next nonzero order in the Taylor series expansion. Neither model fits the data exactly as the BVM for both models at $(\gamma=100\%,\epsilon=0)$ is zero.}
\end{figure}
 \noindent For a single $(\gamma,\epsilon)$ pair, the model's BVM ratio (a prior the models are assumed to be equally likely) is,
\begin{eqnarray}
R(B(\gamma,\epsilon))=\frac{p(A|M_1,D,\gamma,\epsilon)}{p(A|M_2,D,\gamma,\epsilon)},
\end{eqnarray} 
which, because the numerator and denominator is either 0 or 1 in the deterministic case, gives $R(B(\gamma,\epsilon))$ equal to 1, 0, $\infty$, or $\frac{0}{0}$ meaning that the models both agree, model 1 does not agree but model 2 agrees, model 1 agrees but model 2 does not agree, or both models disagree, respectively. Thus, the BVM ratio for a single $(\gamma,\epsilon)$ pair between two deterministic models with completely certain data is not particularly insightful as they either agree or do not agree as defined by $B$. As it may not always be clear precisely what values of $(\gamma,\epsilon)$ one should choose to define agreement, one can meaningfully average (marginalize, analagous to (\ref{marginalizeme})) over a viable volume in the space of $(\gamma,\epsilon)$, with $p(\gamma,\epsilon)=\frac{1}{V}$, and arrive at an averaged Boolean BVM ratio,
\begin{eqnarray}
R(B)=\frac{\sum_{\gamma,\epsilon}p(A|M_1,D,\gamma,\epsilon)}{\sum_{\gamma,\epsilon}p(A|M_2,D,\gamma,\epsilon)}=\frac{N_{1A}}{N_{2A}},
\end{eqnarray}
which is simply a ratio of the number of agreements found for model 1, $N_{1A}$, in the $(\gamma,\epsilon)$ volume to the number of agreements found for model 2, $N_{2A}$, in the selected $(\gamma,\epsilon)$ volume. In our deterministic example, $R(B)= \frac{1108}{2364} = 0.4687$ as model 2 better fits the data, as defined by $B$, for the chosen meaningful $(\gamma,\epsilon)$ volume (which is taken to be the whole tested volume in this toy example). The BVM ratio or the averaged Boolean BVM ratio may be used as a guide for selecting models in the deterministic case given reasonable regions are chosen.

The BVM probability of agreement values as a function of $(\gamma,\epsilon)$ are plotted in Figure \ref{2Non-DeterministicPlot} for model 1 and model 2 in the uncertain model case:
  \begin{figure} [H]
\indent 
  \includegraphics[scale=0.41]{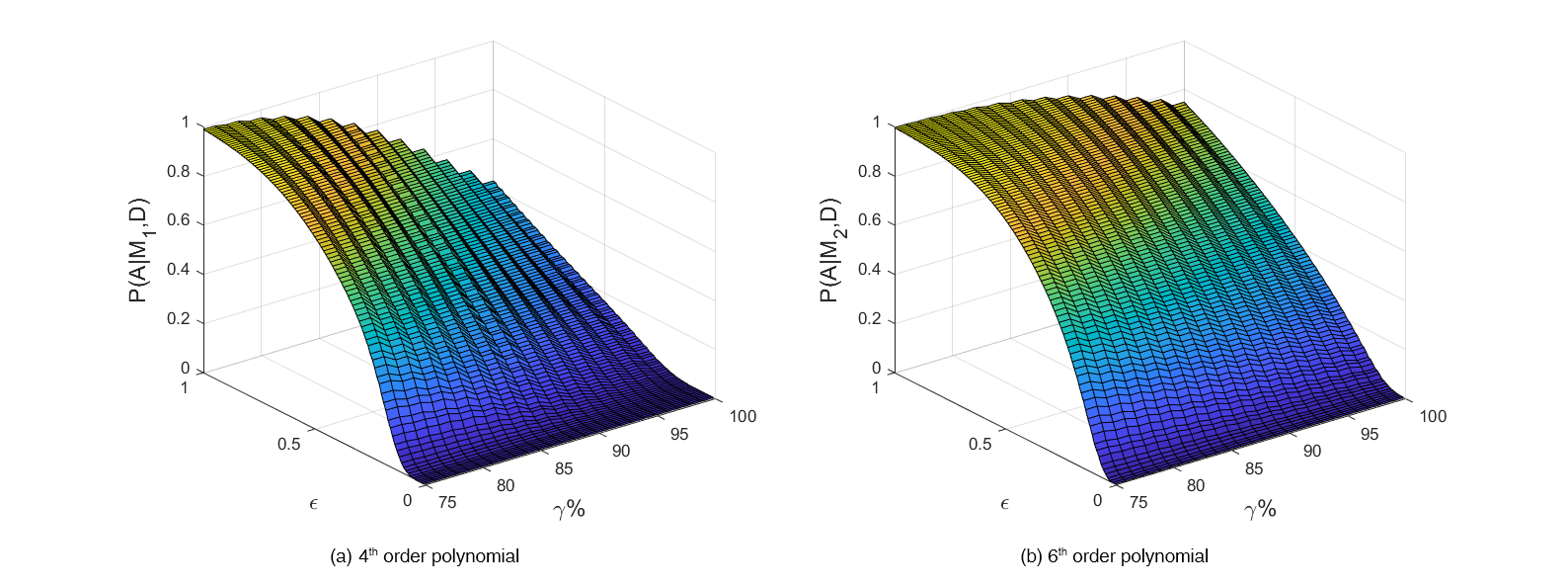}
\caption{\label{2Non-DeterministicPlot} Uncertain Case: The BVM probability of agreement between model 1 and 2 with the data is plotted in the space of $(\gamma,\epsilon)$. The results for model 1 ($4^{\mathtt{th}}$ order polynomial) is plotted on the left and model 2 ($6^{\mathtt{th}}$ order polynomial) is plotted on the right. Because the model paths are uncertain, the BVM probability of agreements for each $(\gamma,\epsilon)$ pair may take any value from zero to one. As expected, model 2 better fits the data in the space of $(\gamma,\epsilon)$ as its BVM is generally larger than that of model 1; however, the BVM values are about equal in cases of large values of $\gamma$ and $\epsilon$ values (the definition of agreement is less stringent and they both ``agree") and in the case of demanding absolute equality ($\epsilon=0$) as neither model fits the data exactly.}
\end{figure}
The BVM ratios $R(B(\gamma,\epsilon))=\frac{p(A|M_1,D,\gamma,\epsilon)}{p(A|M_2,D,\gamma,\epsilon)}$ of the uncertain models are plotted as a function of $(\gamma,\epsilon)$ in Figure \ref{p12}:
  \begin{figure} [H]
\indent \qquad \qquad \qquad \qquad \qquad \:\:
  \includegraphics[scale=0.41]{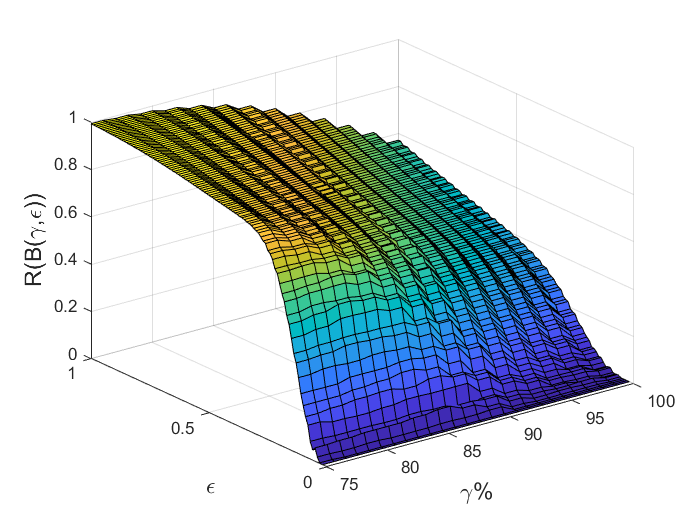}
\caption{\label{p12} This is a plot of the BVM ratios in the uncertain case. Model 2 is generally favored over model 1 as there exist no values greater than one on the plot. The amount the BVM ratio favors model 2 over model 1 decreases as the metric becomes less and less stringent (i.e. as $\gamma$ decreases and $\epsilon$ increases). The $\epsilon=0$ line was removed because neither model agrees with the data exactly. }
\end{figure}
\noindent The averaged Boolean BVM ratio for the uncertain models is,
\begin{eqnarray}
R(B)=\frac{\sum_{\gamma,\epsilon}p(A|M_1,D,\gamma,\epsilon)}{\sum_{\gamma,\epsilon}p(A|M_2,D,\gamma,\epsilon)} = 0.7471,
\end{eqnarray}
which conforms to the notion that model 2 is, generally speaking, the preferable model, and which may be communicated with this single number. We given examples of the BVM ratio correctly selecting models according to abstract and new forms of validation. 

		\section{Conclusion}
 We demonstrated the versatility of the BVM toward expressing and solving model validation problems. The BVM quantifies the probability the model is valid for arbitrary quantifiable definitions of model-data agreement using arbitrary quantified comparison functions of the model-data comparison values. The BVM was shown: to obey all of the desired validation metric criteria \cite{Liu2} (which is a first), to be able to represent all of the standard validation metrics as \emph{special cases}, to supply improvements and generalizations to those special cases, and to be a tool for quantifying the validity of a model in novel model-data contexts. The later was demonstrated by the validation metrics we invented and quantified in our examples. 

Finally, it was shown that one can perform model selection using the BVM ratio. The BVM model testing framework was shown to generalize the Bayesian model testing framework to arbitrary model-data contexts and with reference to arbitrary comparisons and agreement definitions. That is, the BVM ratio may be used to rank models directly in terms of the relevant model-data validation context. The problem of model-data validation may be reduced to the problem of finding/ defining the four BVM inputs: $(\hat{z},z)$, $\rho(\hat{z},z|M,D)$, $f(\hat{z},z)$, and $B(f)$, and computing their BVM value. We find that the BVM is a useful tool for quantifying, expressing, and performing model validation and testing. Our future work involves probabilistically regressing (learning) model parameter distributions that satisfy arbitrary definitions of agreement within the BVM framework.

	\paragraph{Acknowledgments}
	This work was supported by the Center for Complex Engineering Systems (CCES) at King Abdulaziz City for Science and Technology (KACST) and the Massachusetts Institute of Technology (MIT). We would like to thank all of the researchers with the Center for Complex Engineering (CCES), especially Zeyad Al-Awwad, Arwa Alanqari, and Mohammad Alrished. Finally, we would also like to thank Nicholas Carrara and Ariel Caticha.
	\paragraph{References}

	\appendix

	\section{Deriving the other validation metrics from the BVM}
	
	In the following subsections we will show some of the special cases of the Bayesian validation metric. Subsequent improvements or immediate generalizations of the metrics using (\ref{genagree}) are presented when applicable. A detailed review of the majority of these metrics may be found in \cite{Liu2} and the references therein. Table 1 and 2 in Section 4 outline the results.

	\subsection{Reliability metric and probability of agreement}
	There are a few validation metrics related to the reliability metric present in the literature. The reliability metric $r=p(|\expt{\hat{y}}-\mu_{y}|<\epsilon)$ \cite{Rebba} is equal to the probability that the data and the model expectation values are within a tolerance of size $\epsilon$. Their ``probability of agreement" introduced in \cite{Stevens} is closely related to $r$, but instead expresses the quantity as ``the probability the data and the model expectation values \emph{agree} within a tolerance (or sliding tolerance) of $\epsilon$". The reliability metric was expanded in \cite{Sankararaman2} to account for model outputs and data rather than simply comparing the mean of the model prediction against the mean of the data. The improved reliability metric is equal to,
	\begin{eqnarray}
	r_i=\int_{-\infty}^{\infty}\rho(Y|D)\int_{Y-\epsilon(Y)}^{Y+\epsilon(Y)}\rho(\hat{Y}|M)\,d\hat{Y}\,dY,
	\end{eqnarray} 
	where $\rho(\hat{Y}|M)$ and $\rho(Y|D)$ are the full joint probability distributions of the model outputs and the data, respectively. This metric quantifies the probability that the error is less than a value $\epsilon(y)$ on a point to point basis.
	
	
	The BVM is the reliability metric when the comparison values are $\hat{z}=\expt{\hat{y}}$ and $z=\mu_{y}$, and $B$ takes the form of an inequality, being true if $-\epsilon\leq \expt{\hat{y}} -\mu_{y}\leq\epsilon$. If we would like to use a sliding interval of ``tolerance" or ``error acceptance", denote it by $[c_{-}(\mu_{y}),c_{+}(\mu_{y})]$", where $c_{-}<c_+$, and the BVM is,
	\begin{eqnarray}
	p(A|M,D)&=&\int_{\expt{\hat{y}},\mu_{y}} \rho(\expt{\hat{y}}|M) \Theta\Big(c_{-}(\mu_{y})\leq \expt{\hat{y}}\leq c_{+}(\mu_{y})\Big) \,\rho(\mu_{y}|D)\,d \expt{\hat{y}}\,d\mu_{y}\nonumber\\
	&=&\int_{\mu_{y}}\int_{\expt{\hat{y}}=c_{-}(\mu_{y})}^{c_{+}(\mu_{y})}\rho(\expt{\hat{y}}|M)\rho(\mu_{y}|D)\,d\expt{\hat{y}}\,d\mu_{y}=r.\label{19}
	\end{eqnarray}
	This is the reliability metric if $c_{\pm}=\mu_{y}\pm \epsilon$ is a constant and symmetric interval about $\mu_{y}$. 
	
	The BVM is the improved reliability metric when $\hat{z}=\hat{Y}$, $z=Y$, and when the Boolean $B(\hat{z},z)$ is true iff $|\hat{y}-y|\leq\epsilon(y)$ for all $\hat{y},y$ pairs. This is,
	\begin{eqnarray}
	p(A|M,D)=\int_{\hat{Y},Y}\rho(Y|D)\rho(\hat{Y}|M)\Big(\prod_{\hat{y},y\mbox{ \scriptsize{pairs}}}\Theta(|\hat{y}-y|\leq\epsilon(y))\Big)\,d\hat{Y}\,dY=r_i.
	\end{eqnarray} 
	
	The BVM quantifies the probability of square error (or difference) is less than some $\epsilon$ by considering,
	\begin{eqnarray}
	p(A|M,D)=\int_{\hat{Y},Y}\rho(Y|D)\Theta(|\hat{Y}-Y|\leq\epsilon)\rho(\hat{Y}|M)\,d\hat{Y}\,dY.
	\end{eqnarray} 

	Nothing in the BVM requires the variables to be continuous or ordered, so the natural generalization is to let the Boolean expression be true if ``The value $\hat{z}$ is in the subset $S(z)$, which is the set of $\hat{z}$'s agreeing with $z$". For example, if $\hat{z}$'s are strings, $S(z)$ might be the set of words or phrases in $\hat{z}$ that are reasonably synonymous with $z$. This gives the straightforward generalization to accommodate arbitrary data types,
	\begin{eqnarray}
	p(A|M,D)=\sum_{\hat{z},z} p(\hat{z}|M) \Theta\Big(\hat{z}\in S(z)\Big) \,p(z|D)
	=\sum_{z}\sum_{\hat{z}\in S(z)}p(\hat{z}|M)p(z|D),\label{26}
	\end{eqnarray}
	by using sets rather than intervals.


	\subsection{Frequentist validation metric}
	
	To include the frequentist validation metric in the BVM, we will have to express the comparison variables $\hat{z}$ and $z$ and their respective probabilities. The result can be replicated by letting: $z=\mu_{y}$ be the Student-t distribution and $\hat{z}=\expt{\hat{y}}$ have a Dirac delta distribution $\rho(\hat{z}|M,D)=\rho(\expt{\hat{y}}|M,D)=\delta(\expt{\hat{y}}-\expt{\hat{y}}')$ where $\expt{\hat{y}}'$ is the known value of the computational model's expected output.  Because the frequentist validation metric does not force the modeler to define what is meant by \emph{agreement}, we represent this freedom by keeping $B(\hat{z},z)$ general. This gives,
	\begin{eqnarray}
	p(A|M,D)&=& \int_{\hat{z},z} \rho(\hat{z}|M,D)\Theta\Big(B\Big) \,\rho(z|D)\,d\hat{z}\,dz\nonumber\\
	&=& \int_{\expt{\hat{y}},\mu_{y}} \delta(\expt{\hat{y}}-\expt{\hat{y}}')\cdot \Theta\Big(B(\expt{\hat{y}},\mu_{y})\Big) \nonumber\\
	&&\cdot\,\Big(\frac{\Gamma(\frac{\nu+1}{2})}{\sqrt{\nu\pi}\Gamma(\frac{\nu}{2})}\Big(1+\frac{(\mu_{y}-\overline{y})^2}{\nu \overline{s}^2/N}\Big)^{-\frac{\nu+1}{2}}\Big)\,d\expt{\hat{y}}\,d\mu_{y},
	\end{eqnarray}
	where $\overline{y}$ is the population average, $\overline{s}$ is the population standard deviation, and $\nu$ is the degrees of freedom are the parameters of the data's Student-t distribution. Making the coordinate transformations $\expt{\hat{y}}\rightarrow \overline{E}=\expt{\hat{y}}-\overline{y}$ and $\mu_{y}\rightarrow E=\expt{\hat{y}}-\mu_{y}$ gives,
	\begin{eqnarray}
	p(A|M,D)&=& \int_{\overline{E},E} \delta(\overline{E}-\overline{E}')\cdot \Theta\Big(B(\overline{E},E)\Big) \,\cdot\Big(\frac{\Gamma(\frac{\nu+1}{2})}{\sqrt{\nu\pi}\Gamma(\frac{\nu}{2})}\Big(1+\frac{(E-\overline{E})^2}{\nu \overline{s}^2/N}\Big)^{-\frac{\nu+1}{2}}\Big)\,d\overline{E}\,dE\nonumber\\
	&=& \frac{\Gamma(\frac{\nu+1}{2})}{\sqrt{\nu\pi}\Gamma(\frac{\nu}{2})}\int_{E'}\Theta\Big(B(\overline{E}',E')\Big) \,\cdot\Big(1+\frac{(E'-\overline{E}')^2}{\nu \overline{s}^2/N}\Big)^{-\frac{\nu+1}{2}}\,dE',\label{bayesfreq}
	\end{eqnarray}
	with $E'=\expt{\hat{y}}'-\mu_{y}$ and $\overline{E}'=\expt{\hat{y}}'-\overline{y}$. Given that judgments of agreement in the frequentist validation metric are expected to be made based on the confidence level $1-\alpha$ that $E'$ is within the confidence interval, this may be factored into $B(\overline{E}',E')\rightarrow B(\overline{E}',E',\alpha)$, as well as other user defined terms toward expressing agreement. Equation (\ref{bayesfreq}) is thought to be the full BVM's representation of the frequentist validation metric.
	
	
	The immediate generalization to the frequentist validation metric offered by the BVM is to let the model output expectation value $\hat{z}=\expt{\hat{y}}$ have some amount of uncertainty. The uncertainty is perhaps Gaussian or Student $t$ distributed in $\mu_{\expt{\hat{y}}}$ (the true model expectation value) due to only having a finite number of Monte Carlo samples and/or uncertainty induced by discretization error. Because $\hat{z}=\mu_{\expt{\hat{y}}}$ and $z=\mu_{y}$ are both uncertain in general, one generalizes to,
	\begin{eqnarray}
	p(A|M,D,B)= \int_{\mu_{\expt{\hat{y}}},\mu_{y}} \rho(\mu_{\expt{\hat{y}}}|M)\cdot \Theta\Big(B(\mu_{\expt{\hat{y}}},\mu_{y})\Big)\cdot \rho(\mu_{y}|D)\,d\mu_{\expt{\hat{y}}}\,d\mu_{y}.
	\end{eqnarray}
	
	It is interesting to note the consequence of defining a reasonable Boolean expression of agreement on the BVM representation of the frequentist metric. A natural agreement function $B$ for the metric is one that is true if $E_{\mu}=|\mu_{\expt{\hat{y}}}-\mu_{y}|\leq \epsilon$.
	The BVM then gives,
	\begin{eqnarray}
	p(A|M,D)= \int_{\mu_{\expt{\hat{y}}},\mu_{y}} p(\mu_{\expt{\hat{y}}}|M,D)\cdot\Theta\Big(E_{\mu}\leq \epsilon\Big) \cdot p(\mu_{y}|D)\,d\mu_{\expt{\hat{y}}}\,d\mu_{y}\nonumber\\
	= \int_{\mu_{\expt{\hat{y}}}}\int_{\mu_{y}=\mu_{\expt{\hat{y}}}\pm \epsilon} p(\mu_{\expt{\hat{y}}}|M,D) p(\mu_{y}|D)\,d\mu_{\expt{\hat{y}}}\,d\mu_{y}=r.\label{reliabilityfreq}
	\end{eqnarray}
Thus, the frequentist's validation metric is the reliability metric \cite{Rebba} and ``probability of agreement" \cite{Stevens} when reasonable accuracy requirements are imposed on the acceptable difference between the expectation values.

	\subsection{Area and Binned Probability Difference Metric}
	The BVM is able to represent and generalize the area metric by letting the comparison values $(z,\hat{z})$ be the the cdfs in question to be compared $(F(y|D),F(\hat{y}|M))$. The Area metric is,
		\begin{eqnarray}
d[F(\hat{y}|M),F(y|D)]=\int_{-\infty}^{\infty}\Big|F(\hat{y}|M)-F(y=\hat{y}|D)\Big| \,d\hat{y}.
	\end{eqnarray}
The Area metric may be represented by the BVM in a simple way. First allow the comparison value function to be the Area metric functional,
		\begin{eqnarray}
f(z,\hat{z})\equiv d[F(\hat{y}|M),F(y|D)],
	\end{eqnarray}
	and define agreement with a Boolean,
	\begin{eqnarray}
	p(A|M,D)=\Theta\Big(B\Big(d[F(\hat{y}|M),F(y|D)]\Big)\Big).
	\end{eqnarray}
	Here, the cdfs are treated as completely certain ($\delta(\hat{z}-F(\hat{y}|M))$ and $\delta(\hat{z}-F(\hat{y}|M))$ are Dirac delta functionals). The functional form of the Boolean may be decided on given the user's specific validation requirements; however, satisfying some kind of $\epsilon$ threshold $d[F(\hat{y}|M),F(y|D)]\leq\epsilon$ seems logical. Generalizations to the Area metric may be represented analogously with the BVM.  
	
		When the values of the cdfs are uncertain the BVM becomes,
			\begin{eqnarray}
	p(A|M,D)\leftarrow\int \Theta\Big(B(d[F(\hat{y}|M),F(y|D))]\Big)\,
\rho(F(\hat{y}|M),F(y|D))\,dF(\hat{y}|M)\,dF(y|D),\nonumber\\
	\end{eqnarray}
	which may pose computational challenges and require random sampling; however, in some cases it may be permissable to treat the model cdf as known. The extension to uncertain cdf's is usually not considered in the literature; however, the theoretical generalization to the uncertain case is apparent in the BVM framework. As an approximation, one may consider discretizing the Area metric by breaking it into $K$ comparison points,
	\begin{eqnarray}
	d[F(\hat{y}|M),F(y=\hat{y}|D)]\approx \sum_{i=1}^K\Big|F(y_i|M)-F(y=y_i|D)\Big|.
	\end{eqnarray}
	The uncertainty in the cumulative distribution of the data $\rho(z|D)$ is now a finite joint product pdf $\rho(F(y_1|D),...,F(y_K|D)|D)$ over possible cdf values in the $K$ bins, constrained by $F(y_i|D)\leq F(y_{i+1}|D)$. This metric may also be represented with the BVM.

	Alternatively, one may consider a binned probability difference comparison functional,
	\begin{eqnarray}
	d_m[p(\hat{y}|M),p(y=\hat{y}|D)]= \sum_{i=1}^K\Big|p(y_i|M)-p(y=y_i|D)\Big|.
	\end{eqnarray}
	Let
$$
	p(z|D)=P(p_{y_1},...,p_{y_K}|D,\sum_kp_{y_k}=1)
$$
 be the uncertainty for the probability estimate in each data bin due to the random process. Given that the pdf of the model is set to a (presumably known) single pdf function $p(\hat{y}|M)$, the BVM is,
	\begin{eqnarray}
	p(A|M,D)&=&\int_{\hat{z},z} \delta(\hat{z}-\{p(\hat{y}|M)\})\Theta\Big(B\Big(d_m[\hat{z},z]\Big)\Big)\,p(z|D)\,d\hat{z}\,dz\nonumber\\
	&=&\int_{p_{y_1},...,p_{y_K}}\Theta\Big(B\Big(d_m[\{p(\hat{y}_i|M)\},\{p(y_i|D)\}]\Big)\Big)P(p_{y_1},...,p_{y_K}|D)\prod_i(dp_{y_i}).\nonumber\\
	\end{eqnarray}
		Given the form of the distribution $P(p_{y_1},...,p_{y_K}|D)$ is well known (i.e. a Dirichlet distribution after $N$ independent observations of $y$), it can be treated as an expectation value,
	\begin{eqnarray}
	p(A|M,D)=E\Big[\Theta\Big(B\Big(d_m[p(\hat{y}|M),\{p(y_i|D)\})\Big)\Big)\Big]_{\{p(y_i|D)\}},
	\end{eqnarray}
	 and estimated with Monte Carlo. When $K$ is large we may face the curse of dimensionality if the probabilities (bin heights) are themselves uncertain.
	
The number of bins $K$ plays a role similar to $\epsilon$ in that its choice affects the definition/context of agreement represented by the BVM. The binned pdf metric is most informative when $K$ is large because one is checking each region of the pdf for agreement. On the other hand, perhaps when data is limited, there may be instances when having $K=2$ bins is useful, if for instance one was interested in testing a model for representing binary type probabilities (like pass/fail or positive/negative). Using a favorable agreement in this simple binary context to imply that the model works for \emph{otherwise untested} $K>2$ comparisons/contexts/validations, is statistical misrepresentation and should be avoided (see Section \ref{section24}). The BVM requires one to explicitly state the comparison values, the comparison value function, and the definition of agreement to compute $P(A|M,D)$ such that confusion may be avoided and model validation comparisons may be justified. 

Thus, the choice of $K$ ultimately determines the granularity of the definition of agreement being tested. If one uses methods similar to \cite{Knuth} to select $K$, then one is letting the complexity of the data and the number of data points determine the stringency of the definition of agreement. 

	\subsection{Probability Density Function Comparison Metrics}
Another way to gauge the agreement between uncertain data and models is through a pdf comparison metric, $G(\rho_D||\rho_M)$ \cite{Maupin}. Examples of $G(\rho_D||\rho_M)$ are the negative relative entropy or KL divergence, 
\begin{eqnarray}
D(\rho_D||\rho_M)=\int_{y}\rho(y|D)\log\Big(\frac{\rho(y|D)}{\rho(\hat{y}=y|M)}\Big)\,dy,
\end{eqnarray}
the Symmetrized KL divergence $S(\rho_D||\rho_M)=D(\rho_D||\rho_M)+D(\rho_M||\rho_D)$, the Jenson-Shannon divergence $D_{JS}(\rho_D||\rho_M)$, the Hellinger Metric $H(\rho_D||\rho_M)$, the Fisher information distance $\ell(\rho_D||\rho_M)$, and the Wasserstein distance $W(\rho_D||\rho_M)$. This metrics give a notion of ``closeness" between the pdfs that can be used for validation. 

The BVM may represent these validation metrics by letting the comparison value function $f(z,\hat{z})$ be the pdf comparison metric, $$f(z,\hat{z})\equiv G(\rho_D||\rho_M),$$ where $z\equiv \rho_M$ and $\hat{z}\equiv \rho_D$ and are the model and data pdfs, respectively. In the absence of model and data uncertainty (i.e. the pdfs are treated as known functions, e.g. $\rho_D$ is a gaussian with mean=0 and variance=1), the BVM is simply,
$$p(A|M,D)=\Theta\Big(B[f(z,\hat{z})]\Big)=\Theta\Big(B[G(\rho_D||\rho_M)]\Big),$$
which either meets the specifications for them to agree defined by $B$, or does not (e.g. passing a tolerance threshold). Following the structure of the BVM, if there are uncertainties in the functional forms of the pdf's, they may be included into the BVM,
$$p(A|M,D)\leftarrow\int_{\rho_D,\rho_M}\Theta\Big(B[G(\rho_D||\rho_M)]\Big)\rho(\rho_D,\rho_M)\,d\rho_D\,d\rho_{M}.$$
Uncertainties in the data pdfs may come from a lack of data and uncertainties in the model may come from parametric uncertainty (e.g. a Gaussian pdf model $\rho_{M|\mu}$ with an uncertain mean $\rho(\mu)$)\footnote{ One should note that here there is the potential for two different models. One in which $\rho_{M}\equiv \int\rho_{M|\mu}\rho(\mu)\,d\mu$ is marginalized over (in which case the model pdf is ``certain" if integrated analytically) and another in which the model pdf is gaussian $\rho_{M|\mu}$ but there is model parametric uncertainty of the form $\rho(\mu)$.} Similar to the Area metric, these metrics may be discretized from pdf to probability comparison metrics and uncertainty in the pdfs themselves are typically not discussed/quantified in the literature. 

	\subsection{Statistical hypothesis testing\label{stathypsection}}
	
	Normally when statistical hypothesis testing is performed, one constructs the pdf of the relevant test statistic of the data $\rho(z|D)$ and then assumes the null hypothesis is true ($M=D$) counterfactually, which is enforced by setting the ``to be tested population" (the model outputs here) pdf to be equal to the pdf of the test statistic of the data $\rho(\hat{z}|M)\rightarrow \rho(\hat{z}|M=D)$. However, in the present case, we are interested in the general modeling case in which one is able to extract a pdf of the outputs of the model, which we would like to test against data before assuming that they are equal. We will first represent the classical statistical hypothesis test using the BVM and then later supply a version that is more relevant to model validation problems.
	\paragraph{Classical statistical hypothesis testing}
	In classical hypothesis testing, one constructs the pdf of a relevant test statistic $S_{y}\equiv z$ of the data $\rho(z|D)$. The null hypothesis is that the model is equal to the data, which is enforced by setting $\rho(\hat{z}|M)= \rho(\hat{z}|M=D)$. Further, the null hypothesis is not rejected if the test statistic from the model $\hat{z}$ falls within the critical region $[-c_{\alpha},c_{\alpha}]$ that corresponds to the probability $\int_{-c_{\alpha}}^{c_{\alpha}}\rho(z|D)\,dz =1-\alpha$ of the data. The case of not rejecting the null hypothesis is represented by the Boolean expression $B(\hat{z})$, which is true if $-c_{\alpha}\leq \hat{z}\leq c_{\alpha}$ is true -- defining ``agreement" in this case. This results in the following BVM,
	\begin{eqnarray}
	p(A|M=D,D) = \int_{\hat{\hat{z}},z} \rho(\hat{z}|M=D)\Theta\Big(B(\hat{z})\Big) \,\rho(z|D) \,d\hat{z}\,dz\nonumber\\
	= \int_{\hat{z}}\rho(\hat{z}|M=D)\Theta\Big(-c_{\alpha}\leq \hat{z}\leq c_{\alpha}\Big)\,d\hat{z} =1-\alpha,\label{classicalSHT}
	\end{eqnarray}
	because the model was assumed to be equal to the data counterfactually in the null hypothesis and $B(\hat{z},z)=B(\hat{z})$ only. 
	
	The probability of type I error, i.e. rejecting the counterfactually assumed true null hypothesis when it is actually true, is equal to $\alpha$. It should be noted that this is not equal to the probability of finding the model value outside of the data's confidence interval because it is unknown if the null hypothesis (in what results in $\rho(\hat{\hat{z}}|M)\rightarrow \rho(\hat{\hat{z}}|M=D)$) is actually true, or not, because the null hypothesis was merely assumed to be true counterfactually. It should further be noted that with probability $\alpha$ the data's test statistic is outside of its \emph{own} confidence interval. Thus, the probability $\alpha$ both indicates type I error \emph{and} a systematic type error that the wrong sort of comparison is being made, i.e. the wrong Boolean expression was chosen, because, why would we care if the model is within a certain confidence interval if the data is not even within that interval?\footnote{Independent of whether or not the null hypothesis is true*} The probability of type II error, i.e. that the null hypothesis was not rejected when it is actually false, is equal to $\beta_M(\alpha)=1-\int_{-c_{\alpha}}^{c_{\alpha}}\rho(\hat{z}|M)\,d\hat{\hat{z}}$, which in the classical case is difficult to calculate directly because one does not have access to the actual model pdf $p(\hat{z}|M)$ in frequentist probability.
	
	\paragraph{Improved statistical hypothesis testing for validation}
	For the validation cases we are interested in, both $\rho(\hat{z}|M)$ and $\rho(z|D)$ are quantified, and therefore assuming that $\rho(\hat{z}|M)\rightarrow \rho(\hat{z}|M=D)$ would irresponsibly throw away any information sent through the model. We therefore offer the improved statistical hypothesis test for validation using the BVM, which uses both the model and data pdfs. We call this BVM the statistical power BVM.

	For the modified statistical hypothesis test, let the definition of \emph{agreement} be a compound Boolean expression that is true iff both $-c_{\alpha}\leq \hat{z}\leq c_{\alpha}$, that the model test statistic lies in the data's confidence interval, \emph{and} $-c_{\hat{\alpha}}\leq z\leq c_{\hat{\alpha}}$ that the data statistic lies in the model's confidence interval, which corresponds to the probability $\int_{-c_{\hat{\alpha}}}^{c_{\hat{\alpha}}}\rho(\hat{z}|M)\,d\hat{z} =1-\hat{\alpha}$ of the model. By not assuming $M=D$, we remove the possibility that either type I or type II errors can occur; however, systematic errors, that the Boolean expression meant to define agreement is nonsensical, still exist. Thus, while more or less adhering to the type of tests one might perform for model validation using statistical hypothesis testing, we get,
	\begin{eqnarray}
	p(A|M,D)&=&\int_{\hat{\hat{z}},z} \rho(\hat{z}|M,D)\Theta\Big(-c_{\alpha}\leq \hat{z}\leq c_{\alpha}\Big)\Theta\Big(-c_{\hat{\alpha}}\leq z\leq c_{\hat{\alpha}}\Big) \,\rho(z|D)\,d\hat{z}\,dz \nonumber\\
	&=&(1-\beta_M(\alpha))\cdot (1-\beta_D(\hat{\alpha})),\label{improvedSHT}
	\end{eqnarray}
	which is the probability that both the model and the data lie within one another's confidence intervals. The value $1-\beta(\alpha)$ is called the statistical power of the test, but here we have access to both the statistical power of the data and the model. The probability the model and data do not agree as defined by $B$ is given by,
	\begin{eqnarray}
	p(\overline{A}|M,D)=1-p(A|M,D)=\beta_D(\hat{\alpha})+\beta_M(\alpha)-\beta_D(\hat{\alpha})\beta_M(\alpha),
	\end{eqnarray}
	which occurs if either $\hat{z}$, $z$, or both are outside of one another's confidence intervals. The probability for systematic error (that $\hat{z}$, $z$, or both are outside of their own confidence intervals) is equal to $\alpha+\hat{\alpha}-\alpha\hat{\alpha}$; however, there is no conceptual issue with setting $\alpha$, $\hat{\alpha}$, or both equal to 0 as long as both distributions do not span the entire range of possible values (as in such a case the model and data would always agree, which means the test has zero resolving power). Setting $\alpha=\hat{\alpha}=0$ removes the chance of systematic error from the analysis. 
	
	Overall, the use of confidence intervals is suboptimal unless both the model and data pdfs are strictly unimodal. Therefore, rather than using a confidence interval, one may use a ``confidence set", which we define as the smallest set of $\hat{z}$ (as well as for $z$) values that's probability adds to $1-\hat{\alpha}$ (and similarly $1-\alpha$).  Confidence sets are generated by adding the largest probabilities of $\hat{z}$ ($z$) until a confidence level of $1-\hat{\alpha}$ ($1-\alpha$) is met. A confidence interval is only equal to the confidence set if the distribution is unimodal. As the confidence set is the smallest set of values adding up the the confidence level, this set of values is more informative than a confidence interval, which may include many $0$ or low probability events. Using a confidence set improves the resolving power of the metric further. 
	
	The statistical power BVM (\ref{improvedSHT}) is more informative than the classical statistical hypothesis test (\ref{classicalSHT}) because it utilizes the model pdf while also removing type I, II, and (optionally) systematic errors from the test; however its overall resolving power is weak when compared to other metrics. By rewriting the pair of overlapping confidence intervals (or sets) as the set of values in a single ``overlap interval" $I=[c_{\alpha},c_{\alpha}]\cap [c_{\hat{\alpha}}, c_{\hat{\alpha}}]$, one may see that (\ref{improvedSHT}) is a particular case of (\ref{26}) with $S(y)=I$ for $y\in I$ and being the null set otherwise. Thus, the statistical power BVM may be seen as a special case of the generalized reliability metric suggested by the BVM in (\ref{26}) that effectively has large and unvarying tolerance intervals (sets). The use of confidence intervals as well as confidence sets as a means to define agreement effectively coarse grain the probabilities, which removes their informative features. The most stringent definition of agreement between the model and the data leads to Bayesian model testing, which we prove in the next section, as it is indeed equal to the generalized model reliability metric with a zero tolerance for differences $\epsilon=0$.

	\subsection{Bayesian model testing}

	In Bayesian model testing, rather than assuming a particular model is true, one lets the available data determine which model is most likely given that data. Represented probabilistically, out of a set of possible models, Bayesian model testing selects the model with the maximum posterior probability $p(M|Y)$, i.e. the probability of a (previously calibrated \cite{Shankar}) model $M$ given the data set $Y=\{y_i\}$ from data source $D$. As the selection rule for models requires comparing values of $p(M|Y)$, one often constructs the posterior odds ratio,
	\begin{eqnarray}
	R=\frac{p(M|Y)}{p(M'|Y)},
	\end{eqnarray}
	and selects the model $M$ with the highest value of $R$ relative to a chosen base model $M'$. Using Bayes Theorem, the posterior odds ratio may be recast as,
	\begin{eqnarray}
	R=\frac{\rho(Y|M)p(M)}{\rho(Y|M')p(M')}=K\frac{p(M)}{p(M')},
	\end{eqnarray}
	where $K\equiv\frac{\rho(Y|M)}{\rho(Y|M')}$ is known as the Bayes factor, which is the ratio of the likelihoods of the models. The ratio of the prior model probabilities $\frac{p(M)}{p(M')}$ is the ratio of the belief that model $M$ is true prior to examining the data, relative to $M'$. If there is reason to believe that one model is more probable than another due to prior (perhaps statistical) knowledge of the system under investigation, then the Bayesian framework allows one to take this into account through the prior model probabilities. As it is common for the $R$ values to be potentially many orders of magnitude greater than one, the prior model probabilities may be overwhelmed by the Bayes factor. In any case, if there is no reason to prefer one mode over another, one can let the prior model probabilities be equal a-priori. 
	
	Given we are testing some model function $\hat{y}=M(\vec{x},\vec{\alpha})$, we have access to the forward propagation of data and parameters through a given model, the problem of calculating $R$ may be rerouted to the computation of $K$. Using the potentially multidimensional inputs to our models $(\vec{x},\vec{\alpha})$, which represent the input data and the model parameters respectively, the Bayes factor is calculated through forward propagation (marginalization) of these inputs through both models,
	\begin{eqnarray}
	K=\frac{p(Y|M)}{p(Y|M')}=\frac{\int_{\vec{x},\vec{\alpha}}\rho(Y|\vec{x},\vec{\alpha},M)\rho(\vec{x},\vec{\alpha}|M)\,d\vec{x}\,d\vec{\alpha}}{\int_{\vec{x}',\vec{\alpha}'} \rho(Y|\vec{x}',\vec{\alpha}',M')\rho(\vec{x}',\vec{\alpha}'|M')\,d\vec{x}'\,d\vec{\alpha}'}.
	\end{eqnarray}
	The prior probability to the inputs of the model $\rho(x,\alpha|M)$ are the input probability distributions used to propagate uncertainty through the model. The probability $\rho(Y|x,\alpha,M)$ is the probability of the data given by the knowledge of the model function $\hat{y}=M(x,\alpha)$; however, it should be noted that in general the data $Y\neq \hat{Y}=\{y_i\}$,  is not the set of model outputs $\hat{Y}$, as the data $Y$ was collected from the experiment rather than from model outputs. Thus, one must impose the assumptions under which a model is built, i.e. it is built for the purpose of approximating $y_i\approx y_i=M(x_i,\alpha)$, as we will see later. Thus, a more verbose presentation of $K$ is, 
	\begin{eqnarray}
	K=\frac{p(\hat{Y}\equiv Y|M)}{p(\hat{Y}\equiv Y|M')}=\frac{\rho(\hat{Y}\equiv Y|M)}{\rho(\hat{Y}\equiv Y|M')}=\frac{\int_{\vec{x},\vec{\alpha}}\rho(\hat{Y}=Y|\vec{x},\vec{\alpha},M)\rho(\vec{x},\vec{\alpha}|M)\,d\vec{x}\,d\vec{\alpha}}{\int_{\vec{x}',\vec{\alpha}'} \rho(\hat{Y}=Y|\vec{x}',\vec{\alpha}',M')\rho(\vec{x}',\vec{\alpha}'|M')\,d\vec{x}'\,d\vec{\alpha}'},\label{K2}
	\end{eqnarray}
	where $p(\hat{Y}\equiv Y|M)$ is understood to be the sum of the model and the data probabilities that jointly output the same values, i.e. the probability that the any of the possible model and data values turn out to be the same. The form of $\rho(\hat{Y}=Y|\vec{x},\vec{\alpha},M)$ is usually assumed to be something that works (and even better if it is also simple) \cite{Placek}, like the product of Gaussian distributions,
	\begin{eqnarray}
	\rho(\hat{Y}=Y|\vec{x},\vec{\alpha},M)=
	\frac{1}{Z}\exp(-\frac{1}{2\sigma^2}\sum_{i}(M(x_i;\vec{\alpha})-y_i)^2),\label{35}
	\end{eqnarray}
	where $\sigma^2$ is interpreted as the measurement uncertainty of the data.\footnote{The dependence on the data set/ experiment $D$ is therefore implied. } One usually computes the integrals in $K$ using various sampling algorithms such as nested sampling \cite{Sivia,Feroz} or another Markov Chain Monte Carlo technique. As an added bonus, Bayesian model testing has an inbuilt Occam's razor mechanism which penalizes needlessly complex models, i.e. one that would over fit the data by using a large number of uncertain model parameters $\vec{\alpha}$. A clear explanation may be found in \cite{Catichabook}. 

	We can represent Bayesian model testing using the Bayesian validation metric. Let the model comparison value $\hat{z}$ be $\hat{Y}=(y_1,...,y_N)$ that in principle correspond to $z=Y=(y_1,...,y_N)$, a validation set of data. The Boolean expression $B$ is considered to factor into a set of ``and" statements over the individual model output and data points $B(\hat{Y},Y)=B(y_1,y_1)\wedge...\wedge B(y_N,y_N)$, where each $B$ is true iff $y_i=y_i$ exactly. The Bayesian validation metric in this case is then,

	\begin{eqnarray}
	p(A|M,D)&=&\int_{\hat{Y},Y} \rho(\hat{Y}|M,D)\theta(B(\hat{Y},Y)) \,\rho(Y|D)\,d\hat{Y}\,dY\nonumber\\
		&=&\int_{\hat{Y},Y} \rho(\hat{\hat{Y}}|M,D)\delta_{\hat{\hat{Y}},Y} \,\rho(Y|D)\,d\hat{\hat{Y}}\,dY.\label{Bayes1}
	\end{eqnarray}
	In general, computational models may be described by a model function $\hat{Y}=\vec{\hat{y}}=M(\vec{x},\vec{\alpha})$, and given the model pdf was constructed through forward propagation of the uncertainties $\rho(\vec{x},\vec{\alpha})$, the model output pdf is 
	\begin{eqnarray}
	\rho(\hat{Y}|M,D)=\int_{\vec{x},\vec{\alpha}}\rho(\hat{Y}|\vec{x},\vec{\alpha},M,D)\rho(\vec{x},\vec{\alpha})\,d\vec{x}\,d\vec{\alpha}\label{Bayes2}
	\end{eqnarray}
	Substituting (\ref{Bayes2}) into (\ref{Bayes1}), using the trick (\ref{trick}) and (\ref{exactagree}) -- that $$\int_{Y-\epsilon}^{Y+\epsilon} \rho(\hat{Y}|M,D) d\hat{z}\stackrel{\epsilon\rightarrow 0^+}{\longrightarrow} p(\hat{Y}=Y|M,D)=\rho(\hat{Y}=Y|M,D) d\hat{Y},$$ and integrating over $Y$, one finds,
	\begin{eqnarray}
	p(A|M,D)&=&\int_{\vec{x},\vec{\alpha},Y} p(\hat{Y}=Y|\vec{x},\vec{\alpha},M,D)\rho(\vec{x},\vec{\alpha})\rho(Y|D)\,d\vec{x}\,d\vec{\alpha}\,dY\nonumber\\
	&=&\int_{\vec{x},\vec{\alpha}}p(\hat{Y}\equiv Y|\vec{x},\vec{\alpha},M,D)\rho(\vec{x},\vec{\alpha})\,d\vec{x}\,d\vec{\alpha}\nonumber\\
	&=&\Big(\int_{\vec{x},\vec{\alpha}}\rho(\hat{Y}\equiv Y|\vec{x},\vec{\alpha},M,D)\rho(\vec{x},\vec{\alpha})\,d\vec{x}\,d\vec{\alpha}\Big)\,d\hat{Y}\nonumber\\
	&=&\rho(\hat{Y}\equiv Y|M,D)\,d\hat{Y},
	\end{eqnarray}
	which is what is meant by, and is equal to, the probability in the numerator of the Bayes factor (\ref{K2}),
	\begin{eqnarray}
	p(A|M,D)=\rho(\hat{Y}\equiv Y|M,D)\,d\hat{Y}\equiv p(\hat{Y}\equiv Y|M).
	\end{eqnarray}
	That is, we have shown that Bayesian model testing is a special case the generalized model reliability metric in the case of exact agreement (\ref{26}), i.e. $\epsilon=0$, as can be seen by investigating (\ref{Bayes1}). 
	
	Thus, a generalization of Bayesian model testing is to let the definition of agreement have a tolerance $\epsilon>0$ such that the Bayes factor becomes,
	\begin{eqnarray}
	K=\frac{p(A|M,D)}{p(A|M',D)}\longrightarrow K(\epsilon)=\frac{p(A|M,D,\epsilon)}{p(A|M',D,\epsilon)},
	\end{eqnarray}
	where $p(A|M,D,\epsilon)$ is Eq. (\ref{19}). This derivation suggests that the BVM can be used analogously to Bayesian model testing, except with arbitrary definitions of agreement $\epsilon\rightarrow B$, that is, we may construct the BVM factor,
	\begin{eqnarray}
	K(B)=\frac{p(A|M,D,B)}{p(A|M',D,B)}.
	\end{eqnarray}
	Using Bayes Theorem, $p(M|A,D,B)=\frac{p(A|M,D,B)p(M|D,B)}{p(A|D,B)}$, and so we may further construct the BVM ratio,
	\begin{eqnarray}
	R(B)=\frac{p(M|A,D,B)}{p(M'|A,D,B)}=\frac{p(A|M,D,B)p(M|D,B)}{p(A|M',D,B)p(M'|D,B)}=K(B)\frac{p(M|D,B)}{p(M'|D,B)},\label{BVMratioap}
	\end{eqnarray}
	for the purpose of model testing under a general definition of agreement $B$, i.e., we can do model selection under any definition of agreement with the BVM ratio. The ratio $\frac{p(M|D,B)}{p(M'|D,B)}$ is the ratio prior probabilities of $M$ and $M'$, which again, if there is no reason to suspect that one model is a priori more probable than another, one may let $\frac{p(M|D,B)}{p(M'|D,B)}=1$, and then $R(B)\rightarrow K(B)$. With the BVM ratio one could in principle compare data and models with different data types to perform model testing or selection.

\end{document}